\documentclass[aps,pra,reprint,amssymb,superscriptaddress]{revtex4-2} 
\usepackage{amsmath,amssymb,amsfonts}
\usepackage{graphicx}
\usepackage{dcolumn}
\usepackage{bm}
\usepackage[colorlinks=true,citecolor=blue,linkcolor=blue,urlcolor=blue]{hyperref}
\usepackage[mathlines]{lineno}
\usepackage{orcidlink}
\usepackage{float}
\usepackage{subfigure}
\makeatletter
\let\newfloat\newfloat@ltx
\makeatother
\usepackage{algorithm}
\usepackage{algpseudocode}
\usepackage{dsfont}
\usepackage{braket}
\usepackage[dvipsnames]{xcolor}
\usepackage{tabularray}

\definecolor{tms}{rgb}{1,0,0}

\begin{document}

\title{Simulating dynamics of the two-dimensional transverse-field Ising model: a comparative study of large-scale classical numerics}

\author{Joseph Vovrosh\,\orcidlink{0000-0002-1799-2830}}
\email{joseph.vovrosh@pasqal.com}
\let\comma,
\affiliation{PASQAL SAS, 24 rue Emile Baudot - 91120 Palaiseau,  Paris, France}

\author{Sergi Juli\`a-Farr\'e\,\orcidlink{0000-0003-4034-5786}}
\let\comma,
\affiliation{PASQAL SAS, 24 rue Emile Baudot - 91120 Palaiseau,  Paris, France}

\author{Wladislaw Krinitsin\,\orcidlink{0009-0009-2169-1246}}
\let\comma,
\affiliation{Institute of Quantum Control (PGI-8), Forschungszentrum Jülich, D-52425 Jülich, Germany}
\affiliation{Faculty of Informatics and Data Science, University of Regensburg, D-93053 Regensburg, Germany}

\author{Michael Kaicher\,\orcidlink{0000-0001-7986-5127}}
\let\comma,
\affiliation{PASQAL SAS, 24 rue Emile Baudot - 91120 Palaiseau,  Paris, France}

\author{Fergus Hayes\,\orcidlink{0000-0001-7628-3826}}
\let\comma,
\affiliation{PASQAL SAS, 24 rue Emile Baudot - 91120 Palaiseau,  Paris, France}

\author{Emmanuel Gottlob\,\orcidlink{0000-0003-3166-5497}} 
\let\comma,
\affiliation{PASQAL SAS, 24 rue Emile Baudot - 91120 Palaiseau,  Paris, France}

\author{Augustine Kshetrimayum\,\orcidlink{0000-0002-0130-1142}} 
\let\comma,
\affiliation{PASQAL SAS, 24 rue Emile Baudot - 91120 Palaiseau,  Paris, France}

\author{Kemal Bidzhiev} 
\let\comma,
\affiliation{PASQAL SAS, 24 rue Emile Baudot - 91120 Palaiseau,  Paris, France}

\author{Simon B. J\"ager\,\orcidlink{0000-0002-2585-5246}}
\let\comma,
\affiliation{Physikalisches Institut, University of Bonn, Nussallee 12, 53115 Bonn, Germany}

\author{Markus Schmitt\,\orcidlink{0000-0003-2223-8696}}
\let\comma,
\affiliation{Institute of Quantum Control (PGI-8), Forschungszentrum Jülich, D-52425 Jülich, Germany}
\affiliation{Faculty of Informatics and Data Science, University of Regensburg, D-93053 Regensburg, Germany}

\author{Joseph Tindall\,\orcidlink{0000-0003-1335-8637}}
\let\comma,
\affiliation{Center for Computational Quantum Physics, Flatiron Institute, 162 Fifth Avenue, New York,
New York 10010, USA}

\author{Constantin Dalyac\,\orcidlink{0000-0002-0339-6421}}
\let\comma,
\affiliation{PASQAL SAS, 24 rue Emile Baudot - 91120 Palaiseau,  Paris, France}

\author{Tiago Mendes-Santos\, \orcidlink{0000-0001-6827-5260}}
\email{tiago.mendes-santos@pasqal.com}
\let\comma,
\affiliation{PASQAL SAS, 24 rue Emile Baudot - 91120 Palaiseau,  Paris, France}

\author{Alexandre Dauphin\,\orcidlink{0000-0003-4996-2561}}
\let\comma,
\affiliation{PASQAL SAS, 24 rue Emile Baudot - 91120 Palaiseau,  Paris, France}

\begin{abstract}
The quantum dynamics of many-qubit systems is an outstanding problem that has recently driven significant advances in both numerical methods and programmable quantum processing units.
In this work, we employ a comprehensive toolbox of state-of-the-art numerical approaches to classically simulate the  dynamics of the two-dimensional transverse field Ising model. Our methods include three different tensor network techniques---matrix product states, tree-tensor networks, and two-dimensional tensor-networks under the belief propagation approximation---as well as time-dependent variational Monte Carlo with Neural Quantum States. We focus on two paradigmatic dynamical protocols: (i) quantum annealing through a critical point and (ii) post-quench dynamics. Our extensive results show the quantitative predictions of various state-of-the-art numerical methods providing a benchmark for future numerical investigations and experimental studies with the aim to push the limitations on classical and QPUs. In particular, our work connects classical simulability to different regimes associated with quantum dynamics in Rydberg arrays - namely, quasi-adiabatic dynamics, the Kibble-Zurek mechanism, and quantum quenches.
\end{abstract}

\date{\today}

\maketitle

\section{\label{sec:intro}Introduction}

\begin{figure*}
    \centering
    \includegraphics[width=\textwidth]{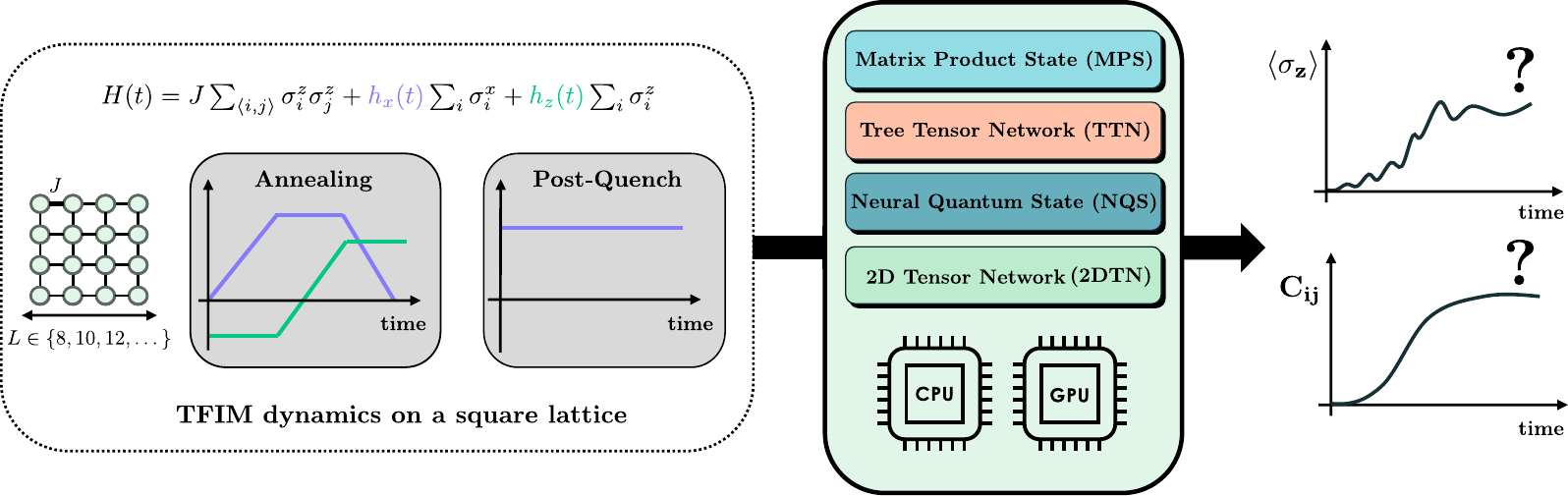}
    \caption{We consider a square layout of atoms and study the Transverse Field Ising model with two types of protocols: annealing and quench. We then use a toolbox of state-of-the-art classical solvers to replicate two types of observables: magnetization and two-point correlations.}
    \label{fig:Fig_1}
\end{figure*}

The last decade has witnessed the rapid development of quantum processing units (QPUs), supported both by academic research and growing industrial involvement. These versatile platforms allow for precise control and manipulation of many-body quantum systems, and therefore offering a new tool for the study of these systems in complex regimes. In particular, a key promise of these devices is their potential to simulate many-body quantum systems in regimes which are inaccessible on classical computers and therefore being used to discover or probe emergent, non-equilibrium phenomena.  
Indeed, in the absence of noise, it has been shown that QPUs can efficiently simulate the dynamics of strongly correlated quantum systems~\cite{feynman_simulating_1982,lloyd_universal_1996}. However, current devices are noisy and, in parallel, unbiased classical numerical methods have been actively developed, giving rise to a stimulating competition pushing the boundaries of classical numerical methods and quantum algorithms for simulations.

QPUs in digital mode have allowed for the large-scale quantum simulation of the kicked Ising model~\cite{kim_evidence_2023} and for the prethermalization study of the digitized Ising model~\cite{haghshenas_digital_2025}. QPUs in analog mode, meanwhile, have opened the door to the large-scale study of non-adiabatic effects near spin-glass transitions~\cite{kim_evidence_2023,king_beyond-classical_2025}, of thermalization and criticality in the one-dimensional XY model~\cite{andersen_thermalization_2025}, and of Haar random distributions in one-dimensional Ising models~\cite{choi_preparing_2023, shaw_benchmarking_2024}. 

While QPU platforms have shown remarkable advances, the limits of classical simulability remain poorly understood, particularly in connection with the diverse dynamical phenomena that can be naturally accessed on quantum devices. In recent years, many works based on classical numerical methods have softened, or even overturned, claims of quantum advantage~\cite{tindall_efficient_2024,mauron_challenging_2025,tindall_dynamics_2025}. Addressing this gap is therefore essential for clarifying the boundary between classical and quantum computational capabilities in the study of many-body dynamics.

Numerous numerical methods have been proposed to address many-body systems. Whereas state-vector simulations are limited by the exponential scaling of their memory requirements with system size, more efficient approaches to simulating quantum dynamics - based on compact variational representations of quantum states, such as tensor networks (TN) and Neural Quantum States (NQS) - have been developed. Tensor network ansätze include, among others, one-dimensional Matrix Product States (MPS)~\cite{stoudenmire_studying_2012}, Tree Tensor Networks (TTN)~\cite{tagliacozzo_simulation_2009, murg_simulating_2010, schuhmacher_hybrid_2025}, and higher-dimensional tensor networks (2D-TN and 3D-TN)~\cite{verstraete_renormalization_2004,jordan_classical_2008,tindall_dynamics_2025}. These architectures can be used to study quantum dynamics by means of tailored algorithms such as the time-dependent variational principle (TDVP)~\cite{haegeman_time-dependent_2011} and the time-evolving block decimation (TEBD)~\cite{vidal_efficient_2004, verstraete_matrix_2004, white_real-time_2004}. In the case of non-tree architectures, the choice of tensor network contraction algorithm becomes important, with choices ranging from highly efficient but approximate BP-based methods ~\cite{jiang_accurate_2008, alkabetz_tensor_2021, tindall_gauging_2023, banuls_matrix_2009,abanin_colloquium_2019, lerose_influence_2021, park_simulating_2025} to more accurate but more expensive approaches such as boundary MPS~\cite{lubasch_algorithms_2014,verstraete_renormalization_2004, guo_block_2023} and the corner transfer matrix renormalization group method~\cite{nishino_corner_1996, orus_simulation_2009}. NQSs are represented by various types of artificial neural networks (ANNs), including restricted Boltzmann machines and convolutional neural networks.
When combined with time-evolution algorithms - such as time-dependent Variational Monte Carlo (tVMC) for NQS - both approaches have successfully simulated the dynamics of relatively large quantum systems, going beyond brute-force state-vector simulations. Finally, recent works tackled successfully quantum dynamics with Pauli propagation methods~\cite{rall_simulation_2019,rudolph_classical_2023,broers_exclusive-or_2024,broers_scalable_2025,begusic_real-time_2025,rudolph_pauli_2025}.

However, it is evident that these numerical approaches are not able to accurately capture real-time dynamics across all regimes, facing limitations in many physically relevant cases. 
For tensor-network methods, this limitation arises from the rapid growth of entanglement, which cannot be faithfully represented with restricted bond dimensions. 
For NQS-based approaches, understanding their limitations remains an open question. 
Numerical evidence suggests that these challenges are not necessarily related to entanglement, but rather to instabilities or sampling complexity in the tVMC algorithm.
The lack of numerical techniques that are valid across all parameter regimes makes a comprehensive and extensive comparison and benchmark of existing numerical techniques a crucial requirement. 

In this work, we focus on many-body dynamics compatible with Rydberg atom QPUs operating in the analog mode. These versatile platforms have several strengths: arbitrary two-dimensional array geometries, continuous-time evolution, dynamical control of system parameters, and long coherence times~\cite{browaeys_many-body_2020, saffman_quantum_2010, henriet_quantum_2020}. 

They therefore allow for the exploration of a wide range of dynamical phenomena. Notable examples considered here include annealing protocols for preparing magnetically ordered states~\cite{scholl_quantum_2021, ebadi_quantum_2021, semeghini_probing_2021}, investigations of non-adiabatic effects associated with universal features~\cite{keesling_quantum_2019} and coarsening dynamics~\cite{manovitz_quantum_2025}, post-quench protocols for studying quantum thermalization~\cite{shaw_benchmarking_2024, choi_preparing_2023, shaw_experimental_2025, darbha_probing_2025}, and protocols to study phenomena related to high-energy physics~\cite{vovrosh_meson_2025, mark_observation_2025, gonzalez-cuadra_observation_2025}.

To assess the capabilities of classical numerics and find challenging regimes compatible with Rydberg atom QPUs in a two-dimensional square array, we first consider a cross-benchmark of a toolbox of numerical approaches to simulate both annealing and post-quench dynamics; see Fig.~\ref{fig:Fig_1}.
Specifically, we employ three tensor-network methods - MPS, TTN, and 2DTN - as well as tVMC with NQSs. In an attempt to compare these methods equally, we restrict ourselves to equal hardware, namely, access to A100 Nvidia GPUs.
This analysis is complemented by more conventional investigations of the errors associated with each individual method, followed by a proposed framework to verify expected symmetry constraints of observables during the dynamics. We further verify that mean-field and semiclassical methods cannot capture the behavior described by the previous methods---an indication of strong correlation of the studied post-quench dynamics.

The main contribution of this work is to delineate the capabilities of state-of-the-art numerical methods in describing different dynamical phenomena that are realizable in Rydberg array QPUs. Our results show that although the considered classical numerical methods are complementary in describing annealing protocols, these approaches struggle to capture post-quench dynamics in certain parameter regimes.

The rest of this paper is organized as follows. In Sec.~\ref{sec2}, we introduce the effective model used to describe the dynamics of Rydberg atom arrays. Sec.~\ref{sec3} provides a summary of the numerical methods employed in this work, with further details presented in the Appendix. In Sec.~\ref{sec4}, we perform a cross-benchmark of the different methods for two dynamical protocols: (i) annealing and (ii) post-quenches. Secs.~\ref{sec5} and \ref{sec6} discuss the convergence criteria for each numerical method.
To aid the reader, we give a summary of all results in \ref{sec7_v2}.
We then address how Ising physics can be probed utilizing am array of Rydberg atoms in Sec~\ref{sec7}. Finally, Sec.~\ref{sec8} summarizes our main conclusions.

\section{ \label{sec2} Model and choice of the quantum dynamics}

We consider the two-dimensional quantum transverse-field Ising Hamiltonian (2D TFIM) on a square lattice array
\begin{equation}
H= J\sum_{\langle i,j \rangle}\sigma^z_i\sigma^z_j + h_{x}(t) \sum_i \sigma^x_i + h_{z}(t) \sum_{i}  \sigma^z_i,
\label{eq:H}
\end{equation}
where $J$ represents the energy scale of the nearest-neighbor interactions, and $h_{x}(t)$ and $h_{z}(t)$ are the time-dependent transverse and longitudinal fields, respectively. 
Throughout this work, we consider square lattices with open boundary conditions and $N = L \times L$ qubits, where $L$ is the linear system size. 
Furthermore, consistent with experimental realizations, we focus on the TFIM with antiferromagnetic interactions, i.e., $J>0$.

At equilibrium, the 2D TFIM is one of the simplest models that exhibits both ground-state and finite-temperature phase transitions between paramagnetic and magnetically ordered phases. 
As we discuss below, this property gives rise to various dynamical phenomena that have been explored using  Rydberg atom arrays. 
Particularly, in such devices (and as considered in this work), the dynamics of interest starts from an initial fully polarized product state,  
\begin{equation}
\ket{\psi(0)} = \ket{\downarrow \downarrow \cdots \downarrow},
  \label{initState}
\end{equation}
which is time-evolved according to
\begin{equation}
\ket{\psi(t)} =  {\cal T} \left[ e^{-i   \int_0^{t}  H(t^{\prime}) d t^{\prime}} \ket{\psi(0)} \right].
\label{eq:time_evol}
\end{equation}
The time-dependent pulse shapes $h_{x}(t)$ and $h_{z}(t)$ can then be chosen such that the system follows either (i) an annealing or (ii) a post-quench dynamics, as illustrated in Fig.~\ref{fig:Fig_1}.

\begin{figure}[t]
    \includegraphics[scale=1]{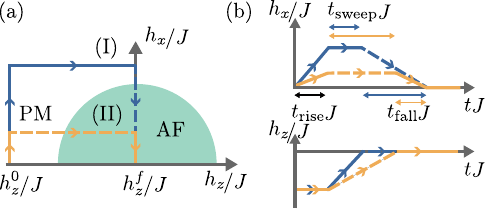}
   \caption{Panel (a) shows a schematic phase diagram for the \textit{antiferromagnetic} transverse field Ising model on the square lattice, while panel (b) displays the sweep profiles characterized by three time scales: $t_{\mathrm{rise}}$, $t_{\mathrm{sweep}}$, and $t_{\mathrm{fall}}$. The corresponding trajectories in the phase diagram for the two annealing schedules discussed in Sec.~\ref{sec4.1} are shown as blue dotted and red dashed lines, which we denote as annealing (I) and (II), respectively.  
 In the two cases, we consider that the initial value for the longitudinal field is $h^0_z = -8J$, while the one at the end of the sweep is $h^f_z = 0$. Furthermore, we consider for (I) $t_{\mathrm{rise}}J=1.5$, $t_{\mathrm{sweep}}J=1.5$, and different values of  $t_{\mathrm{fall}}$ and for (II)  $t_{\mathrm{rise}}J=1.5$, $t_{\mathrm{fall}}J=1.5$, and different values of  $t_{\mathrm{sweep}}$; more details about the annealing sweeps are presented in the main text.
 }
    \label{fig2}
\end{figure}

\subsection{Quantum Annealing}
\label{sec2.1}

In the quantum annealing protocol, the initial state evolves under a  time-dependent Hamiltonian. Particularly, we consider a parameter sweep driving the system from a region of the ground-state phase diagram hosting a paramagnetic (PM) phase to a region of antiferromagnetic (AFM) phase (see Fig.~\ref{fig2}).

Such an annealing protocol has recently been explored both experimentally with Rydberg atoms~\cite{scholl_quantum_2021,ebadi_quantum_2021} and through numerical approaches~\cite{schmitt_quantum_2022,mendes-santos_wave-function_2024}.  
A key feature is that during the sweep, the system crosses a second-order quantum critical point with a lowest energy gap $\Delta$ that can vanish at the crossing point, a generic hallmark of continuous quantum phase transitions.
For example, in the experimentally relevant regime of finite system size, the gap scales as $\Delta \sim L^{-z}$ near the critical point (where $L$ represents the linear size of the array and $z =1$ is the dynamical critical exponent at the transition point)~\cite{sachdev_quantum_2011, laumann_quantum_2015}.

The dynamics of the annealing can then be understood as follows. 
Far from the critical point, the dynamics adiabatically follows the ground state of the time-dependent Hamiltonian.
In the vicinity of the crossing point, when the finite-size gap is larger than the typical sweep rate of the parameters (in Fig.~\ref{fig2}, this is characterized by $1/t_{\text{sweep}}$ or $1/t_{\text{fall}}$), the dynamics can still be considered quasi-adiabatic, which allows one to prepare a final time state with the ground-state order.
In fact, such protocols are employed to perform state preparation characterized by magnetic order~\cite{browaeys_many-body_2020}.

Conversely, when the typical sweep rate is larger than the finite-size gap, a breakdown of adiabaticity occurs. In particular, close to the crossing point, such non-adiabatic effects can be understood in terms of generalizations of early studies of the dynamics of topological defects in classical systems, as proposed by Kibble~\cite{kibble_topology_1976} and Zurek~\cite{zurek_cosmological_1985}, which is known as the quantum Kibble-Zurek mechanism~\cite{zurek_dynamics_2005} and manifests itself in physical observables exhibiting universal scaling behavior~\cite{schmitt_quantum_2022}. These non-adiabatic effects can also suppress long-range order at the end of the sweep.

As illustrated in Fig.~\ref{fig2}, in  Sec.~\ref{sec4} we consider two reference sweeps, which are representative of (I) a quasi-adiabatic regime and (II) an annealing process exhibiting non-adiabatic effects, in particular the quantum Kibble-Zurek mechanism, see App.~\ref{sec:qkz}.

\begin{figure}[t]
    \includegraphics{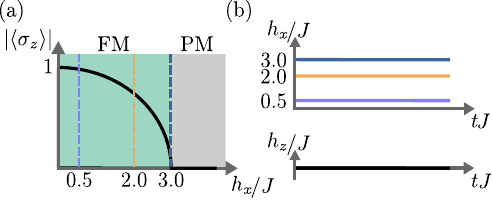}
   \caption{Panel (a) shows a schematic of the zero-temperature \textit{ferromagnetic} phase diagram of the transverse-field Ising model, $H_\text{FM}$, on the square lattice for $h_z/J = 0$. We highlight the three values of the transverse field $h_x/J$ considered for post-quench dynamics in this work: one at, and two below, the critical field $h_c/J \approx 3$. Panel (b) displays the three quench protocols employed in our analysis, corresponding to the field values of $h_x/J$ indicated in panel (a) and $h_z/J = 0$.
 }
    \label{fig3}
\end{figure}

\subsection{Post-quench dynamics}
\label{sec2.2}

In the post-quench dynamics, after a rapid change of the field, the initial state evolves under the influence of a time-independent Hamiltonian. Particularly, we focus on post-quench dynamics with $h_{z} = 0$ and finite values of $h_{x}$. 

Owing to time-reversal symmetry and the fact that the initial state [Eq.~\eqref{initState}] is described by a real-valued wave function, the post-quench dynamics in the antiferromagnetic TFIM are equivalent to those following a quench to the ferromagnetic TFIM~\cite{frerot_multi-speed_2018}, i.e.,
\begin{equation}
    H_\text{FM} = -H.
\end{equation}
In particular, since our initial state corresponds to the ground state of the ferromagnetic TFIM with $h_x/J = 0$, it is convenient to analyze the quench dynamics within this ferromagnetic picture. This viewpoint provides a clearer physical intuition for the evolution of the system and allows us to interpret the post-quench behavior more transparently as a function of the ratio $h_x / J$, see Fig.~\ref{fig3}.

Recently, several numerical studies have explored the post-quench dynamics of the 2D ferromagnetic TFIM~\cite{khasseh_discrete_2020,balducci_localization_2022,tindall_confinement_2024,pavesic_constrained_2025,krinitsin_roughening_2025,krinitsin_time_2025}. 
The regime of strong interactions (i.e., $h_x/J \leq 1$) can be efficiently simulated using tensor network approaches, such as TTN~\cite{pavesic_constrained_2025,krinitsin_roughening_2025,krinitsin_time_2025} and 2DTN~\cite{tindall_confinement_2024}. 
 Interestingly, these approaches have enabled the investigation of various dynamical phenomena in the 2D TFIM. 
Examples include the mechanism of Hilbert space fragmentation (HSF)~\cite{yoshinaga_emergence_2022,hart_hilbert_2022}, the emergence of confinement of excitations~\cite{balducci_localization_2022,tindall_confinement_2024,pavesic_constrained_2025}, and the connection between dynamics of interfaces and the roughening transition occurring at equilibrium~\cite{krinitsin_roughening_2025,krinitsin_time_2025}.

The regime of intermediate values of $h_x$, however, is much less explored, due to limitations of numerical methods, as we discuss in Sec.~\ref{sec4}.
In this regime, the long-time behavior of local observables is expected to be governed by the thermal states of the ferromagnetic TFIM, allowing the equilibrium phase diagram to serve as a guide for identifying physically relevant dynamical regimes~\cite{mondaini_eigenstate_2016, mondaini_eigenstate_2017}. 

In particular, in Sec.~\ref{sec4}, we focus on the post-quench dynamics in the vicinity of the ground-state quantum phase transition (QPT), which occurs at $h_x^{c}/J \approx 3.04$~\cite{jordan_classical_2008}, see Fig.~\ref{fig3}. This non perturbative regime can exhibit non-trivial dynamics related to the QPT. 

Furthermore, the dynamical phase transition (DPT), which is related to the equilibrium thermal transition, is another interesting regime. The DPT is characterized by the asymptotic behavior of the time-averaged order parameter $\langle \sigma^z \rangle$, which either remains finite or vanishes at long times for an initially fully polarized state.
These two long-time behaviors, which depend on $h_x/J$, correspond to ordered and disordered dynamical phases, respectively~\cite{zunkovic_dynamical_2018,khasseh_discrete_2020}. The value of the quench parameter that separates these phases defines the DPT. For the 2D TFIM, this critical value is expected at $h_t^{*}/J \approx 1.8$~\cite{blas_test_2016,suzuki_quantum_1987, suzuki_quantum_2013}.

\subsection{Physical Observables}
\label{sec2.3}

We consider experimentally accessible observables to characterize the capabilities of the numerical approaches in describing the dynamical phenomena described above. These observables include single-particle quantities, such as the local magnetization
\begin{equation}
m_i =  \left< \sigma^z_{r_i} \right>,
\end{equation}
where ${r_i}$ represents the position of a site of the square lattice, and two-body correlation functions
\begin{equation}
    C(\delta_i) = \left< \sigma^z_{\vec{r_0}} \sigma^z_{\vec{r_0} -\vec{\delta}_i}\right> - \left< \sigma^z_{\vec{r_0}} \right> \left< \sigma^z_{\vec{r_0} -\vec{\delta}_i} \right>,
    \label{eq:corre}
 \end{equation}
where $\delta_i$ represents the Euclidean distance to a reference site $\vec{r_0}$.
The latter quantity also allows us to probe magnetic long-range order and universal scaling behavior associated with the quantum Kibble–Zurek mechanism.

Before proceeding, it is worth emphasizing that the TFIM serves as an approximate effective description of Rydberg atom arrays, enabling us to place all numerical methods on equal footing while still capturing the essential features of the different dynamical regimes under consideration. While this work mainly focuses on the dynamics of the TFIM model, in Sec.~\ref{sec7}, we discuss  experimental effects, such as long-range interactions, inhomogeneities in local longitudinal fields and comparisons of the time scales we access with simulations (which throughout this paper is given in units of [$1/J$]) with the experimental time scale in Rydberg arrays QPUs.  

\section{\label{sec3} Brief review of the classical simulation toolbox for dynamics}

Recently, significant efforts have been made to advance the capabilities of  tensor-network and time-dependent variational Monte Carlo (tVMC) approaches for simulating quantum dynamics in 2D arrays. 
In this section, we briefly review recent results and challenges associated with these methods, with a particular focus on simulations of the dynamics of the 2D TFIM. We also describe technical aspects of the hyperparameters used to obtain the results presented in the following sections.

\subsection{Tensor network approaches}

Tensor network approaches are among the most used and successful techniques to describe quantum many-body systems in non-perturbative limits. The strength of these methods lies in an efficient encoding of the entanglement structure in the tensor-network description of the quantum state, which can be considered almost exact in some scenarios where entanglement area laws~\cite{hastings_area_2007, eisert_colloquium_2010}, or tree-like spread of correlations can be exploited. While initially conceived for studying ground-states of 1D systems with MPS~\cite{white_density-matrix_1993, schollwock_density-matrix_2011}, the technical developments of the field in the last decades have led to algorithms to study unitary time evolutions~\cite{white_real-time_2004, secular_parallel_2020, paeckel_time-evolution_2019, haegeman_time-dependent_2011, vidal_efficient_2004, verstraete_matrix_2004}, and the introduction of more general architectures such as TTN and 2DTN~\cite{verstraete_renormalization_2004, jordan_classical_2008}, also referred as 2D projected-entangled-pair-states (PEPS). Notably, the tensor-network toolbox has been recently leveraged to efficiently and accurately simulate ~\cite{tindall_efficient_2024, tindall_confinement_2024, begusic_fast_2023,liao_simulation_2023} some of the most recent large-scale experiments on qubit-based QPUs~\cite{king_beyond-classical_2025, kim_evidence_2023}, and to investigate other dynamics scenarios in square lattices up to $16 \times 16$ qubits~\cite{pavesic_constrained_2025} and $24 \times 24$~\cite{pavesic_scattering_2025}.

Here we briefly discuss some key technical aspects of the three tensor network approaches used in this work, namely MPS and TTN ansätze, where dynamics are implemented in a time-dependent variational principle, and a 2DTN architecture, where gate-based dynamics are simulated with the help of the BP approximation~\cite{jiang_accurate_2008, alkabetz_tensor_2021, tindall_gauging_2023}, whilst expectation values are measured with a more accurate boundary MPS contraction scheme~\cite{lubasch_algorithms_2014,verstraete_renormalization_2004, rudolph_simulating_2025}.

\subsubsection{MPS-TDVP  approach}

The MPS simulations in this work are carried out using the Python package \texttt{emu-mps}~\cite{bidzhiev_efficient_2025}. 
Time evolution is implemented through the two-site TDVP algorithm. 
We define the MPS on a snaking path across the square lattice: beginning from the bottom-left corner, the path proceeds rightward along the first row and then continues from the left of the second row. 
All simulations are performed with the largest available bond dimension for each lattice size, chosen in multiples of 100, and executed on an NVIDIA A100 GPU with 40~GB of memory. 
The time step for the evolution is set to $\Delta t \sim 0.01/J$, ensuring sufficient resolution of the dynamics. 

\subsubsection{TTN-TDVP approach}

The TTN simulations are performed using the Julia package \texttt{TTN.jl}~\cite{tausendpfund_ttnjl_2024}. 
For the time evolution we implement the one-site TDVP algorithm, applied to a binary tree structure following the approach of Ref.~\cite{krinitsin_time_2025}. 
As in the MPS case, we employ the largest available bond dimension for each lattice size, here chosen in multiples of 20, compatible with the memory of a 40~GB NVIDIA A100 GPU. The real-time dynamics are simulated using a time step of $\Delta t = 0.01/J$.

\subsubsection{2DTN-BP approach}

In this approach, a 2DTN representation of the quantum state is used whose structure reflects the underlying 2D square lattice geometry. The amount of entanglement is controlled by the dimension of the virtual bonds of the tensor network $\chi_{\text{2D}}$. To evolve the state in time we use a Trotterized gate evolution of the Hamiltonian using a small time step $\Delta t_{\text{2D}}$.  To truncate the state following the application of a two-site gate, a truncated SVD is performed conditioned on incident tensors which form a rank-1 approximation of the environment. These can be obtained efficiently via the belief propagation algorithm~\cite{jiang_accurate_2008, alkabetz_tensor_2021, tindall_gauging_2023}. Such an approximation is exact for tree tensor networks, but is also often reasonable in lattices with small loops when the correlation length is small or only a small number of singular values are discarded in the SVD. Finally, to compute local and nonlocal expectation values at a given evolution time $t$, we first truncate the 2DTN state down to $\chi_{\text{2D}}'<\chi_{\text{2D}}$ using BP-conditioned SVDs, and then use a boundary MPS contraction scheme of the network, with MPS's of dimension $r_{\text{2D}}$, to compute the observables~\cite{lubasch_algorithms_2014, verstraete_renormalization_2004}. In the present work, we implement this approach by means of the \texttt{TensorNetworkQuantumSimulator.jl} package~\cite{tindall_tensornetworkquantumsimulatorjl_2025}, an open-source \textit{Julia} package for quantum simulation with tensor networks of generic topology built on top of the \texttt{ITensors.jl}~\cite{fishman_itensor_2022} library. Unless otherwise specified, we fix: $\chi_{\text{2D}}=40$, $\chi_{\text{2D}}'=24$, $r_{\text{2D}}=32$, and $\Delta t_{\text{2D}}J=0.01$. The Trotterized time evolutions are executed on an AMD EPYC 9654 CPU, using a maximum peak RAM of $\lesssim 50$ GB. The computation of observables using boundary MPS are performed on a 40~GB NVIDIA A100 GPU.

\subsection{NQS-tVMC approach}
Time-dependent variational Monte Carlo (tVMC) has emerged as a promising alternative to tensor-network approaches for simulating dynamics in two-dimensional arrays. These methods can be implemented with a wide range of ansätze, ranging from physically inspired wave functions to artificial neural networks, also referred to as NQSs.
An appealing aspect of these variational states is that their expressive power is not necessarily constrained by entanglement and may, in principle, circumvent both the exponential scaling of exact methods and the entanglement limitations of tensor networks~\cite{carleo_localization_2012,carleo_solving_2017,levine_quantum_2019,sharir_neural_2022}. Previous results have already demonstrated their success in specific physical regimes. Examples include the verification of the quantum Kibble–Zurek mechanism in 2D~\cite{schmitt_quantum_2022}, the quasi-adiabatic dynamical preparation of many-body states~\cite{carleo_simulating_2024,mendes-santos_wave-function_2024,mauron_predicting_2025} and simulations of the dynamical structure factor in the vicinity of quantum critical points~\cite{mendes-santos_highly_2023}.

However, these approaches encounter limitations and shortcomings in several physically relevant regimes~\cite{schmitt_simulating_2025}. Understanding  these issues remains an active area of research, with previous studies indicating numerical instabilities associated with solving the equations of motion~\cite{schmitt_quantum_2020}, as well as difficulties related to the complexity of Monte Carlo sampling in certain regimes~\cite{sinibaldi_unbiasing_2023}. The precise origins of these limitations are still not fully understood. As a consequence, it is often difficult to determine a priori whether the tVMC approach will succeed or fail in simulating a given dynamics. In this work, we investigate the capabilities of tVMC (in particular, the NQS approach) by performing cross-benchmarks with tensor-network methods across different dynamical regimes relevant to Rydberg array experiments.

\subsubsection{Methodological aspects}

We consider NQS based on CNN~\cite{schmitt_quantum_2020}. In particular, we consider a two-layer CNN with $\alpha$ and $\alpha -1 $ channels in the two consecutive layers, with a filter with linear size $F = L/2$ and with 
suited zero-padding of the computational basis configuration to account for the open boundary
conditions. We also consider symmetry projections - including spatial reflection,
rotation - are applied during the simulation.
In the present work, we implement this approach by means of the open-source \textit{jVMC}~\cite{schmitt_jvmc_2022} library.

Furthermore, an important technical detail is that we consider simulations in a local basis that differs from the computational basis of the Hamiltonian~\eqref{eq:H}.  
This aspect is fundamental, as product states aligned with the computational basis, such as the initial state considered here, are pathological for the TDVP algorithm.
In addition to overcome this issue, we observe that the choice of specific bases can improve the results for the dynamics, as briefly discussed in App.~\ref{app:nqs}.

In all NQS-tVMC simulations, we restrict our hardware access to NVIDIA A100 GPUs with 40~GB of RAM, applying parallelization where appropriate.

\section{ \label{sec4} Cross-benchmarking of classical numerics}

In this section, we present results of simulations of the quantum annealing and post-quench dynamics discussed in Sec.~\ref{sec2}. Particularly, we focus on system sizes and time scales where no established ground-truth results exist. We thus base the evaluation of the reliability of the numerical outcomes mainly on cross-benchmarks of the various methods discussed in the previous section. In addition to presenting the time-evolution of the magnetization, $\langle \sigma_{r_i}^z \rangle$, and the connected correlations, $C(\delta_i)$, we perform quantitative comparisons between results of different approaches at representative times of the dynamics. 

We define the discrepancy between the results of two approaches, $A_1$ and $A_2$, through
\begin{equation}
\epsilon_{z} = \frac{ 2 \sum_i \left| \langle \sigma_{r_i}^z \rangle_{A_1} - \langle \sigma_{r_i}^z \rangle_{A_2} \right|}{L}.
\end{equation}
Similarly, for the connected correlations, we consider
\begin{equation}
\epsilon_{zz} =  \frac{\sum_{i} \left| C(\delta_i)_{A_1} - C(\delta_i)_{A_2} \right|}{\sum_{i} C(\delta_i)_{A_1}}  ,
\end{equation}
where the sum over the indices $i$ is taken along the subset of sites lying on the same horizontal line as a center-most site.

\subsection{Quantum Annealing}
\label{sec4.1}

We start by considering cross-benchmarks of the two representative annealing sweeps illustrated in Fig.~\ref{fig2}. 

We anticipate that these two parameter regimes correspond to distinct physical scenarios, distinguished by how AFM correlations develop during the parameter sweep. In particular, in annealing (I), the finite-size gap near the crossing point prevents non-adiabatic effects from destroying the AFM order characteristic of the ground-state Hamiltonian at the end of the sweep. Conversely, in annealing (II), non-adiabatic effects suppress the AFM order in the final state. In the following two subsections, we cross-benchmark the numerical approaches in these two regimes and then provide a more detailed discussion of non-adiabatic effects and the quantum Kibble–Zurek mechanism in Subsection~\ref{Sec.4.1.3}.

\begin{figure*}[t]
    \includegraphics{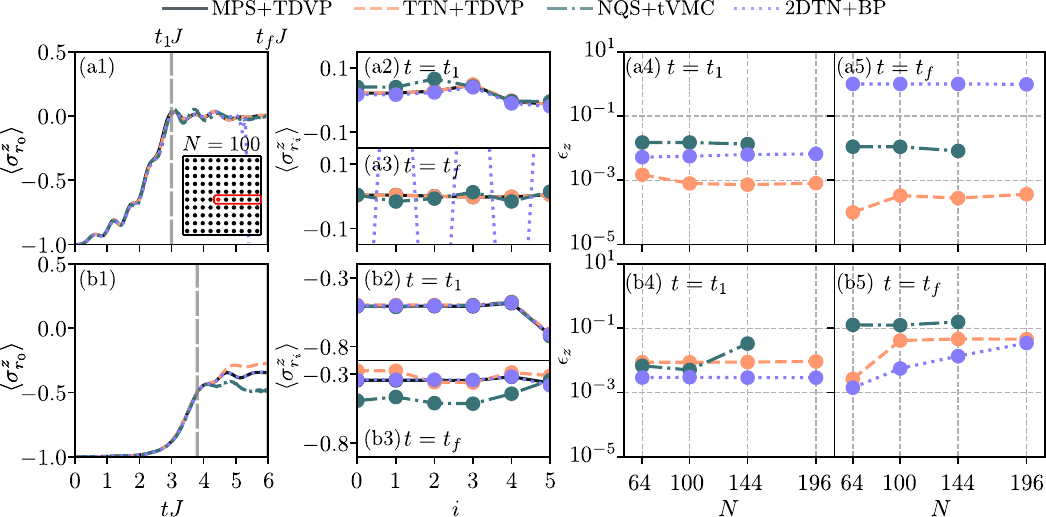}
   \caption{\textbf{Annealing (magnetization).} Cross-benchmarks of local magnetization, $\langle \sigma_{r_i}^z \rangle$, obtained with different numerical methods (MPS, TTN, NQS and 2DTN) during the two annealing sweeps described in Fig.~\ref{fig2}. The set of panels labeled (a) and (b) correspond to sweeps (I) and (II), respectively. In panels (a1) and (b1), we present the time dependence of $\langle \sigma_{r_0}^z \rangle$. Panels (a2) and (a3) [(b2) and (b3)] show the spatial distribution of $\langle \sigma_{r_i}^z \rangle$ along a horizontal line at two representative times, $t_1$ and $t_f$. Panels (a4) and (a5) [(b4) and (b5)] display $\epsilon_z$ between each method and MPS results at $t_1$ [$t_f$] for different system sizes.}
    \label{fig4}
\end{figure*}

\begin{figure*}[t]
    \includegraphics{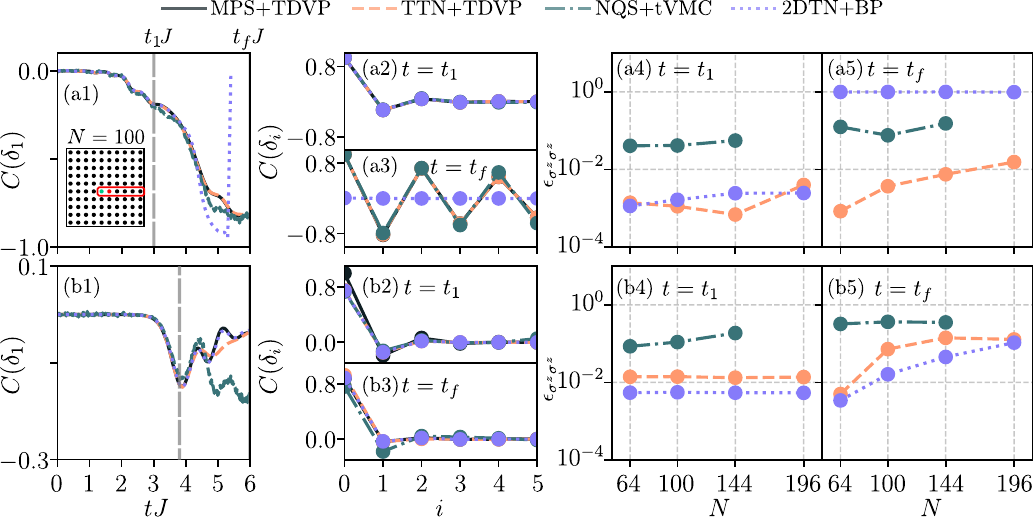}
   \caption{\textbf{Annealing (correlations).} Cross-benchmarks of connected correlations, $C(\delta_i)$, obtained with different numerical methods (MPS, TTN, NQS and 2DTN) during the two annealing sweeps described in Fig.~\ref{fig2}. The set of panels labeled (a) and (b) correspond to sweeps (I) and (II), respectively. In panels (a1) and (b1), we present the time-dependence of nearest-neighbor correlations, $C(\delta_1)$. Panels (a2) and (a3) [(b2) and (b3)] show the spatial distribution of $C(\delta_i)$ along a horizontal line at two representative times, $t_1$ and $t_f$. Finally, panels (a4) and (a5) [(b4) and (b5)] display the discrepancies between the results obtained by the different methods, $\epsilon_{zz}$, at $t_1$ and $t_f$ in relation to MPS results for different system sizes. }
    \label{fig5}
\end{figure*}

\begin{figure*}[t]
    \includegraphics{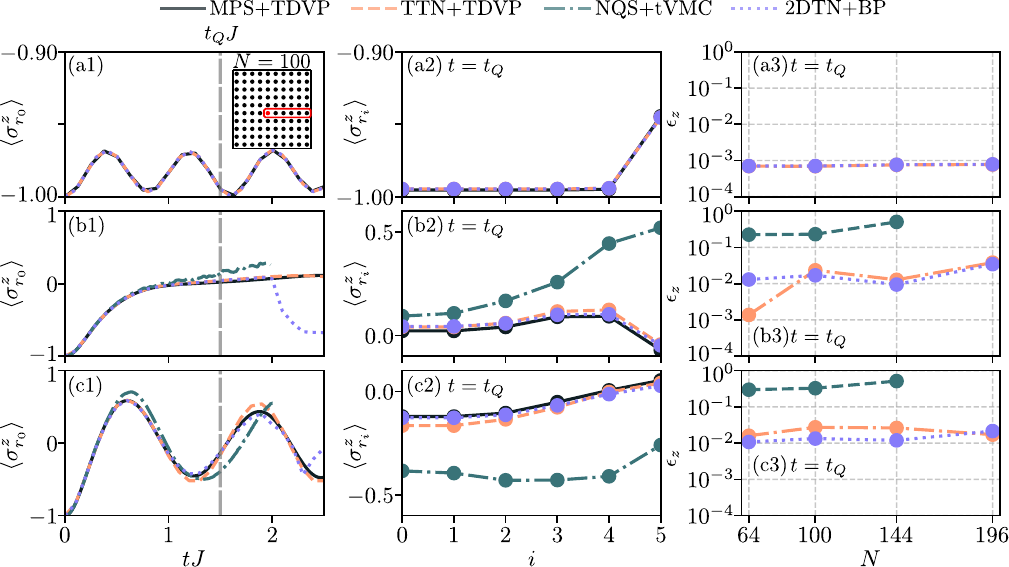}
    \caption{\textbf{Post quench (magnetization).} Cross-benchmarks of local magnetization, $\langle \sigma_{r_i}^z \rangle$, obtained with different numerical methods (MPS, TTN, NQS and 2DTN) for post-quench dynamics. The set of panels (a), (b) and (c) correspond to quenches for $h_x/J=0.5$, $h_x/J=2.0$ and $h_x/J=3.0$, respectively. In panels (a1), (b1) and (c1), we present the time dependence of $\langle \sigma_{r_0}^z \rangle$. Panels (a2), (b2) and (c2)  show the spatial distribution of $\langle \sigma_{r_i}^z \rangle$ along a horizontal line at the representative time $t_Q$. Finally, panels (a3), (b3) and (c3) display the discrepancies between the results obtained by the different methods, $\epsilon_z$, at $t_Q$  in relation to MPS results for different system sizes.}
    \label{fig6}
\end{figure*}

\begin{figure*}[t]
    \includegraphics{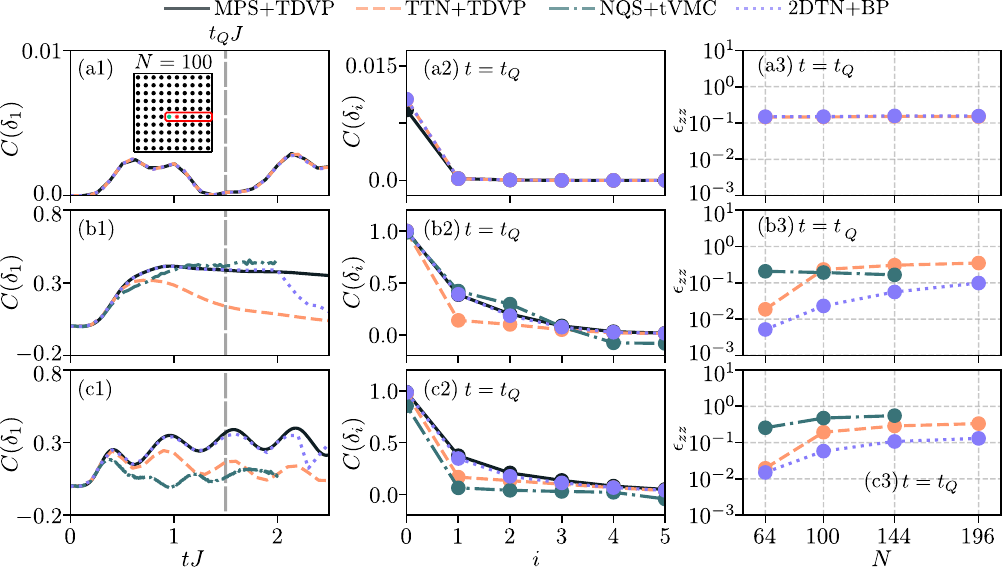}
    \caption{\textbf{Post quench (correlations).} Cross-benchmarks of correlations, $C(\delta_i)$, obtained with different numerical methods (MPS, TTN, NQS and 2DTN) for post-quench dynamics. The set of panels (a), (b) and (c) correspond to quenches for $h_x/J=0.5$, $h_x/J=2.0$ and $h_x/J=3.0$, respectively. In panels (a1), (b1) and (c1), we present the time dependence of nearest-neighbor correlations, $C(\delta_1)$. Panels (a2), (b2) and (c2)  show the spatial distribution of $C(\delta_i)$ along a horizontal line at the representative time $t_Q$. Finally, panels (a3), (b3) and (c3) display the discrepancies between the results obtained by the different methods, $\epsilon_{zz}$, at $t_Q$ in relation to MPS results for different system sizes.}
    \label{fig7}
\end{figure*}

\subsubsection{Results: Annealing I}

In the annealing sweep (I)  (see Fig.~\ref{fig2}), the system enters the antiferromagnetic dome from the top by tuning the transverse field $h_x(t)$, while keeping $h_z/J = 0$; the maximum value of the transverse field in this case is $h_x^m/J=3.5$.  
The rate of this sweep is determined by $t_{\mathrm{fall}}$.  
Here, we focus on cross-benchmarking classical numerical methods for an annealing protocol with total time $t_f J = 6.0$ and $t_{\mathrm{fall}} J = 3.0$.  We note that the crossing point to the antiferromagnetic phase occur at   $t_1J \approx 3.2$,
Additional results for other sweep rates are presented in App.~\ref{sec:qkz}.  

We first compute the residual energy, i.e. the difference between the instantaneous and ground-state energies, at the end of the sweep and we observe that the latter remains low in this case.

Results of the time evolution of the magnetization $\left< \sigma^z_{r_0} \right>$ and the nearest-neighbor connected correlation $C(\delta_1)$, where $\delta_1$ represents a vector connecting the reference site $r_0$ to its nearest-neighbor site,
are presented in  panels (a1) of Figs.~\ref{fig4} and~\ref{fig5}, respectively, for a system with $N = 100$.
We note that results of MPS, TTN and NQS qualitatively agree throughout the whole annealing. Results of  2DTNs, however, deviate from the other approaches at larger times. 

More precisely, we observe that all methods properly describe the build-up of nearest-neighbor AFM correlations appearing at $tJ \approx 3.0$, see panel (a1) of Fig.~\ref{fig5}. This is consistent with the proximity of the crossing point to the ordered phase, occurring at $tJ = 3.2$. To further inspect the onset of AFM long-range order for $tJ \ge 3.2$, we consider the spatial distributions of the local magnetization, $\left< \sigma^z_{r_i} \right>$, and connected correlations, $C(\delta_i)$, along a horizontal line of the square lattice; see panels (a2-a3) of Figs.~\ref{fig4} and~\ref{fig5}. In particular, for the final-time state, we observe that MPS, TTN, and NQS describe the long-range AFM correlations, while 2DTN fails to capture such behavior; see panel (a3) of Fig.~\ref{fig5}. The latter can be explained by the fact that the BP approximation is poor when the correlation length is very large and the loops in the tensor network are small. It is worth mentioning, however, this can sometimes be overcome if BP is replaced by more accurate contraction algorithms when applying gates. Whilst this comes at much higher computational expense, it has been frequently and effectively employed in ground-state studies~\cite{jordan_classical_2008, yang_efficient_2025} of two-dimensional models.

We also assess the quantitative differences between the results of different approaches by considering the discrepancies of the magnetization 
$\epsilon_z$ and relative correlations $\epsilon_{zz}$ in relation to MPS results. We choose MPS as the benchmark, not because it is likely to be the ground truth in these regimes, but because it is the most established and widely used method as well as known to be exact for systems of low entanglement. In particular, we focus on 
two specific times:  $t_1J \approx 3.0$, and another at final time, $t_fJ = 6.0$.
Panels (a4-a5) of Fig.~\ref{fig4} show that $\epsilon_z$ does not exceed $10^{-2}$ for the MPS-NQS and MPS-TTN cross-benchmarks; for  MPS-2DTN the results give $\epsilon_z \approx 10^{0}$ for all considered system sizes  at the final time, although MPS-2DTN still agrees with a great precision with other approaches for $t=t_1$. Similar trends are also observed for the correlations as illustrated in Fig.~\ref{fig5} (a4-a5).

The NQS results for $N = 196$ exhibit convergence difficulties arising from the initialization of the state in the rotated basis (see App.~\ref{app:nqs}). While alternative NQS architectures may mitigate this problem, addressing it lies beyond the scope of this work. We therefore restrict our analysis for NQS to $N \le 144$.

 \subsubsection{Results: Annealing II}

In our second example, we consider an annealing sweep in which the system enters the AFM dome from the side by tuning the longitudinal field, $h_x(t)$, while the transverse field is kept constant at the value $h_x^m/J = 0.5$; see Fig.~\ref{fig2}.
In this case, the rate of sweep is characterized by $t_{\mathrm{sweep}}$. 
As before,  we focus on cross-benchmarks of classical numerics in an annealing with total time $t_f J = 6.0$.

Results of the time evolution of magnetization and nearest-neighbor correlations are presented in  panels (b1) of Figs.~\ref{fig4} and~\ref{fig5}, respectively, for a system with $N = 100$. In contrast to the quasi-adiabatic case,  MPS and 2DTN qualitatively match throughout the whole sweep, while TTN and NQS deviate from those results. 

More specifically, we note that all methods describe a build-up of nearest-neighbor AFM correlations at $tJ \approx 4.0$.
A further inspection of the onset of order at the end of the sweep is made in panels (b2-b3) of Figs.~\ref{fig4} and~\ref{fig5}. Different from the Annealing (I), all methods show that the correlations decay to zero with the distance $\delta_i$, which is a clear evidence that the quantum state at $t_f$ does not exhibit long-range AFM order. As we discuss in more detail below, non-adiabatic effects play an important role in this case, which explains the suppression of magnetic order at the final time.

Although all the methods seem to agree qualitatively in regard to physical properties of the annealing, we observe a clear quantitative discrepancy in the results.
To characterize such discrepancies, we consider  $\epsilon_z$ and  $\epsilon_{zz}$ at $t_1J \approx 3.8$ and $t_fJ = 6.0$ in panels (b4-b5)  of Figs.~\ref{fig4} and~\ref{fig5}, respectively.
For the magnetization, we observe that $\epsilon_z \approx 0.1$ at the end of the sweep for MPS-TTN and MPS-NQS, while for MPS-2DTN  $\epsilon_z < 0.01$, signaling a great agreement between these approaches.

It is worth anticipating that, as will be discussed in Sec.~\ref{sec5} and~\ref{sec6}, 2DTN and MPS provide ground truth results in this regime for $L=10$. However, we observe that for TTN and NQS, we cannot obtain convergence with those results for the variational state used here. When increasing the system size beyond $L=10$, we observe that the error between MPS and 2DTN increases considerably. We expect that in such large systems 2DTN is the most accurate method, since this ansätze and the BP approximation capture efficiently short-range correlated states in 2D. On the other hand, while MPS seems to provide a ground truth at $L=10$, the required bond dimension for fixed error is expected to grow exponentially with the system size due to the MPS 1D mapping of the 2D lattice. In contrast, for a 2DTN ansatz, the required bond dimension for a finite error at a given time should saturate with system size.

\subsubsection{Quantum Kibble-Zurek mechanism} 
\label{Sec.4.1.3}

Before closing this section, we clarify some aspects related to the physical properties of Annealing (I) and (II). In particular, we probe the QKZ mechanism by assessing the expected universal scaling of $C(\delta_i)$ for parameters in the vicinity of the crossing points; see App.~\ref{sec:qkz} (in particular, Fig.~\ref{fig_sm_qkz}). Our results show that for annealing (II), the values of $C(\delta_i)$ for different sweep rates can be well described by a universal scaling function, while for annealing (I) we do not observe such a collapse of the results. This behavior is consistent with the scenario in which, due to finite-size, the dynamics is quasi-adiabatic  for Annealing (I), while for (II), non-adiabatic effects plays an important role and, as a result, the properties of observables in the vicinity of the crossing point are described by the QKZ mechanism~\cite{schmitt_quantum_2022}.

\subsection{Post-quench dynamics}

We now consider post-quench dynamics for $h_z / J = 0$. As discussed in Sec.~\ref{sec2.2}, the time evolution of the fully polarized initial state is equivalent to the dynamics described by a ferromagnetic Hamiltonian (i.e., $H_\text{FM} = -H$, where $H$ is defined in Eq.~\ref{eq:H}).

As a guide to our analyses, it is worth commenting on some expected results for the magnetization, $\langle \sigma^z \rangle$. In the regime of strong interactions, $h_x / J \ll 1$, the dynamics of the initial state, which is described using perturbation theory, couples only to the zero-momentum superposition in the single spin-flip subspace. This leads to approximately two-level dynamics, where the  $\langle \sigma^z \rangle$ oscillates with a monochromatic angular frequency related to the ferromagnetic gap, $\Omega_{\text{FM}} \approx 4J$.
Conversely, in the opposite regime of weak interactions, i.e., $h_x / J \gg 1$, the dynamics are characterized by Rabi oscillations. In this case, $\langle \sigma^z \rangle$ oscillates between its two extremal values with a monochromatic angular frequency $\Omega_{R} \approx 2h_x$.

Below, we consider three post-quench evolutions to $h_x / J = 0.5$, $2.0$, and $3.0$ (with $h_z / J = 0.0$ in all cases), as illustrated in Fig.~\ref{fig3}. The last two cases correspond to regimes in the vicinity of a dynamical phase transition and the quantum critical point of the 2D TFIM, respectively, where the perturbative arguments described above are no longer valid.

\subsubsection{Results}

Results of the time evolution of magnetization, $\left< \sigma^z_{r_i} \right>$, and nearest-neighbor connected correlations $C(\delta_i)$ obtained with different numerical methods are presented in Figs.~\ref{fig6} and~\ref{fig7}.

For $h_x/J = 0.5$, the dynamics is characterized by persistent small oscillations of the $\left< \sigma^z_{r_i} \right>$ and $C(\delta_1)$ up to the largest time scales considered in the simulations, which is consistent with the perturbative regime of strong interactions discussed earlier, see Figures~\ref{fig6}~(a1) and~\ref{fig7}~(a1). In this case, the results of all methods exhibit good agreement up to the maximum time considered for the dynamics~\footnote{We omit the NQS results in Fig.~\ref{fig7} (a1--c1) for the correlations, since the Monte Carlo error bars are larger than the $y$-scale used in the graph.}. We point out that approximate methods are also able to properly describe the dynamics, which highlights the low amount of entanglement involved. 

Away from the perturbative regime, the dynamics is  characterized by damping of the oscillations of $\left< \sigma^z \right>$.
For $h_x/J=3.0$, and times $tJ < 1.5$, for example, there is a qualitative agreement between the methods. In particular, $\left< \sigma^z_{r_i} \right>$ oscillates around zero with a period of  oscillation of $ T J \approx 1$, and corresponding angular frequency $\Omega \approx 2 \pi J $, which is close to the Rabi frequency, $\Omega_{R} = 2.0 h_x$, expected in the regime of weak interactions, i.e., $h_x / J \gg 1$.
For $h_x/J=2.0$, we observe that the results of $\left< \sigma^z_{r_i} \right>$ for all methods tend to a stationary value close to zero, suggesting that the system thermalizes.

However, for the correlations, we observe appreciable deviations between the results of all methods beyond a certain time, which hinder the understanding of the physical properties of the dynamics for longer times. In particular, for both $h_x/J=2.0$ and $h_x/J=3.0$, we observe that all results deviate for times $tJ > 2.0$ when $L \ge 10$; see Figs.~\ref{fig6}~(b1--c1) and~\ref{fig7}~(b1--c1).

A more nuanced discussion of the results for $tJ < 2.0$ is provided in panels~\ref{fig6}~(b2--c2) and~\ref{fig7}~(b2--c2). 
For example, although we observe good quantitative agreement in the local magnetization between MPS and TTN results, we clearly see discrepancies in the nearest-neighbor correlations for times $tJ < 1$. Such discrepancies are also evident in the decay of correlations at $t_Q J = 1.5$. We also observe larger discrepancies between NQS and MPS for $tJ > 1.0$; this is also evident in the distribution of magnetization and in the decay of correlations presented in Figs.~\ref{fig6} and~\ref{fig7}~(b2--c2). A further discussion of these cases is provided in App.~\ref{app:nqs}. Finally, when comparing 2DTN and MPS, we observe quantitative agreement both in the local magnetization and correlations for $tJ < 1$, which seem to indicate that both methods provide relatively well converged results in this regime at $L=10$. Eventually, for $tJ\approx 1.5$, we start to see a discrepancy between the 2DTN and MPS results. Such discrepancy increases with the system size, as shown in panels (b3) and (c3) of Figs.~\ref{fig6} and~\ref{fig7}. Since we expect a faster scaling of the error with the system size in the MPS method, due to the 1D mapping of the 2D lattice, we predict that 2DTN could be the most accurate method to simulate early-time post-quench dynamics in this system. Nevertheless, even for such method it seems very challenging to accurately describe the dynamics beyond $tJ\approx 2$, as shown in Figs.~\ref{fig6}-\ref{fig7}. More details on the scaling of the accuracy of the 2DTN method with computational resources is provided in App.~\ref{app:2dtn}. 

Note that App.~\ref{app:results} presents the full time dynamics of the bulk observables across all three regimes and all system sizes studied. Furthermore, in App.~\ref{app:runtimes} we detail the run-times, and corresponding scaling, required to obtain these results.

\paragraph*{Mean-field and semiclassical approaches} To provide a comprehensive benchmark we also employ efficient simulation techniques that are based on spin mean-field (SMF)~\cite{weiss_hypothese_1907}, parity violating fermionic Gaussian states (PV-FGS)~\cite{KaicherInPrep}, and simulations based on the discrete truncated Wigner approximation (TW)~\cite{schachenmayer_many-body_2015}. Since we are here mostly focusing on the capabilities of these methods to describe the emergent dynamics we have moved the technical description of these methods to App.~\ref{app:approx_sol}. We remark at this point that while SMF disregards all true spin-spin correlations $C(\delta_i)=0$, the FGS and TW methods capture certain correlations due to e.g. the non-local spin-fermion mapping or the use of sampled semiclassical trajectories~\cite{hackl_lucas_fabian_aspects_2018,schachenmayer_many-body_2015,perlin_spin_2020,hosseinabadi_user-friendly_2025}, respectively. The comparison of the performance of these methods to MPS is visible in Fig.~\ref{fig:mf_sc_res} grouped for different parameter regimes (a),(d) $h_x/J=0.5$; (b),(e) $h_x/J=2$; (c),(f) $h_x/J=3.0$; and mean-field quantities $\langle\sigma^z_{r_0}\rangle$ (a),(b),(c); and connected correlation functions $C(\delta_1)$ (d),(e),(f). As mentioned above, SMF is unable to predict any non-vanishing  value of $C$. We find that PV-FGS is able to reproduce the initial dynamics of the MPS in the perturbative regime $h_x/J=0.5$. Complementary to that regime, TW is able to reproduce the MPS dynamics in the regime $h_x /J = 3.0$ at least on short to intermediate timescales. Most notably, all three methods fail to describe the dynamics at a point in time well before the more strongly correlated methods such as MPS start becoming unreliable. This particular example but also the partial failure of the considered mean-field and semiclassical methods in certain regimes highlight that a reliable description of strong correlations is needed for a correct unraveling of the dynamics.

\section{ \label{sec5} Convergence of different numerical results}

To complement our cross-benchmark analysis described above, we discuss a more standard approach to verify the convergence of the different numerical results. 
For this purpose, we consider metrics that quantify the accumulated error of the various time-dependent algorithms used to evolve tensor-network or NQS ans\"atze. Such errors, which is inherent to the approximations (or truncations) necessary to formulate the time-dependent algorithms is referred as the \textit{projection error} here.

While some aspects of this discussion have been addressed previously, additional considerations are introduced in this paper. 
We therefore provide the relevant details regarding the definition of the projection error for the different approaches. 
Readers who are not interested in the technical aspects of the methods may wish to proceed directly to Subsec.~\ref{sec5_e}, which contains a summary of the results.

\subsection{Matrix Product States}
In TDVP-based time evolution, the MPS tensors are updated locally while maintaining a fixed maximum bond dimension \(\chi_{\text{max}}\). After each local evolution step, the updated tensor is decomposed via singular value decomposition (SVD), and the singular values are truncated to preserve this bond-dimension constraint. The discarded singular values introduce a local truncation error, which quantifies the deviation of the evolved tensor from the exact result within the variational manifold.

Formally, if \(M = U S V^\dagger\) denotes the local tensor after an update, and only the largest \(\chi_{\text{max}}\) singular values \(\{ s_i \}_{i=1}^{\chi_{\text{max}}}\) are retained, the local truncation error is given by
\begin{equation}
\epsilon_{\text{loc}}^2 = \sum_{i > \chi_{\text{max}}} s_i^2 .
\end{equation}
This represents the loss of norm due to truncation at that update step. Over a full TDVP sweep, the cumulative truncation error for a given time step can be estimated by summing over all local contributions,
\begin{equation}\label{eq:mps_error}
\epsilon_{\text{MPS}} = \sum_{\text{updates}} \epsilon_{\text{loc}}^2 ,
\end{equation}
which provides a practical measure of the accumulated deviation from the exact time evolution within the fixed bond-dimension manifold. Monitoring \(\epsilon_{\text{MPS}}\) thus serves as an indicator of simulation accuracy and convergence~\cite{paeckel_time-evolution_2019, kloss_time-dependent_2018}.

\subsection{2D Tensor Networks}

In the belief propagation method the dynamics are simulated by applying one and two-qubit gates to the 2DTN via a Trotterization of the  evolution operator. When applying a two-qubit gate \textit{exactly} to a state with a bond dimension $\chi_{2D}$, $\ket{\Psi_i(\chi_{2D})}$, the resulting state $\ket{\Psi_{i+1}(\chi_{2D}')}$ has a bond dimension $\chi_{2D}'$ that is enlarged by a multiplicative factor $r$ equal to the \textit{rank} of the two-site gate. Thus it often needs to be truncated back down to $\ket{\Psi_{i+1}(\chi_{2D})}$ to keep the calculation manageable when sequentially applying large numbers of gates.

This is done via a singular value decomposition (SVD) conditioned on rank-1 environment tensors found via belief-propagation. The error made via this truncation can be approximated as the sum of the square of the discarded singular values~\cite{rudolph_simulating_2025}, i.e. 

\begin{equation}
\epsilon_{\text{BP},i}^{\text{gate}}\equiv \sum_{j=\chi_{2D}+1}^{\chi_{2D}'} \sigma^2_{\text{BP}, j}\approx 1-|\bra{\Psi_{i+1}(\chi_{2D})}G_i\ket{\Psi_{i+1}(\chi_{2D})}|^2.
\end{equation}
where, $G_i$ is a given gate and we assumed that the tensors are normalized such that $\sum_{j=1}^{\chi_{2D}} \sigma^2_{\text{BP}, j}=1$. Note that this approximation is exact only in the case the virtual bonds of $\ket{\Psi_{i+1}}$ do not form loops. If the Trotter step has $m$ gates, we can then approximate the average gate infidelity in a Trotter step as~\cite{zhou_what_2020, rudolph_simulating_2025}
\begin{equation}
    \epsilon_{\text{BP}}\approx 1-\left(\prod_m 1-\epsilon_{\text{BP},i}^{\text{gate}}\right)^{\frac{1}{m}}.
\end{equation}

As a final remark, here we focus on the gate error for the 2DTN-BP approach, which is assumed to be the dominant source of error for the dynamics considered in this work. However, note that even in the case $\epsilon^{\text{trotter}}_\text{BP}\approx 0$, one should also consider the errors associated with the Trotterization of the dynamics using a small but finite $\Delta t_{2D}$ and the fact that the computation of observables at a final time involves a truncation of the state to $\chi_\text{2D}'\leq \chi_\text{2D}$, followed by a boundary MPS contraction with $r_\text{2D}$.

\subsection{Tree Tensor Network and Neural Quantum State approaches}
 
Finally,  we consider a metric that quantify  the error accumulated during the TDVP algorithm (also denoted as TDVP error) to quantify errors of the TTN and tVMC approaches used in this work.

It is worth stressing that this approach is applicable to all variational methods formulated as follows:
given a quantum state $\ket{\psi_{\vec{\theta}(t)}}$, where the time-dependent vector $\vec{\theta}(t)$
represents the parameters of the variational state, the TDVP minimizes the objective function
\begin{equation}
s^2(t) = {\cal{D}}(\ket{\psi_{\vec{\theta}(t + \delta t)}},
e^{-i\hat H \delta t}\ket{\psi_{\vec{\theta}(t)}})^2\ ,
\label{eq:tdvp_err}
\end{equation}
where ${\cal{D}}$ denotes the distance measured by the Fubini–Study metric.

An expansion up to second order in the timestep $\delta t$ yields the infinitesimal formulation of the distance measure.
For the NQS-tVMC approach by doing such expansion one obtains (see e.g. Ref.~\cite{schmitt_simulating_2025}) 
\begin{equation}
\epsilon_{\text{NQS}} = ds^2 = dt^2\Big(\dot{\vec\theta}^T\mathbf{S}\dot{\vec\theta}-2\text{Re}\big[\vec F\big]^T\dot{\vec\theta}+\text{Var}_{\psi_\theta}[\hat H]\Big) \label{eq:tdvp_error_tvmc},
\end{equation}
where $\mathbf{S}$ is the quantum Fisher matrix and and $\vec{F}$ the so-called force vector (see App.~\ref{app:nqs}). Similarly, one can define for  TTN
\begin{align}
\epsilon_{\text{TTN}} = ds^2 = dt^2\Big(\text{Var}_{\psi_\theta}[\hat P \hat H] &- 2\ \big[\braket{\hat H \hat P \hat H} - E \braket{\hat P \hat H}\big] \nonumber \\  
&+ \text{Var}_{\psi_\theta}[\hat H]\Big) \label{eq:tdvp_error_ttn},
\end{align}
see App.~\ref{app:tdvp_err} for a derivation.
These quantities evaluate by how much the time-evolved variational state deviates from an exactly evolved one at each timestep.

For the NQS-tVMC, we also consider the accumulated error up to a maximum time
\begin{equation}
\tilde{R^2}(t)=\int_0^t\sqrt{ds^2}=\int_0^t\sqrt{\epsilon_{\text{NQS}}}\ .
\label{eq:projerror_tdvp}
\end{equation}
Note, that this quantity does not represent the actual, exact error of the time evolution, instead it acts as an upper bound.

It is worth noticing that
\begin{equation}
r^2(t) = \frac{{\cal{D}}(\ket{\psi_{\vec{\theta}(t + \delta t)}}, 
e^{-i\hat H \delta t}\ket{\psi_{\vec{\theta}(t)}})^2}
{{\cal{D}}(\ket{\psi_{\vec{\theta}(t + \delta t)}},\ket{\psi_{\vec{\theta}(t)}})^2},
\label{eq:old_tdvp_err}
\end{equation}
has been used as an alternative error measure used in the literature~\cite{carleo_solving_2017, schmitt_quantum_2020}. This measure of the TDVP error can, however, be biased by physical properties of the dynamics, because ${\cal{D}}(\ket{\psi_{\vec{\theta}(t + \delta t)}},\ket{\psi_{\vec{\theta}(t)}})^2=\text{Var}_{\psi_\theta}[\hat H]$. While the denominator is intended to serve as a natural scale to quantify the error, it becomes deceptive in cases of small or vanishing variance.

\begin{figure*}
    \centering
    \includegraphics{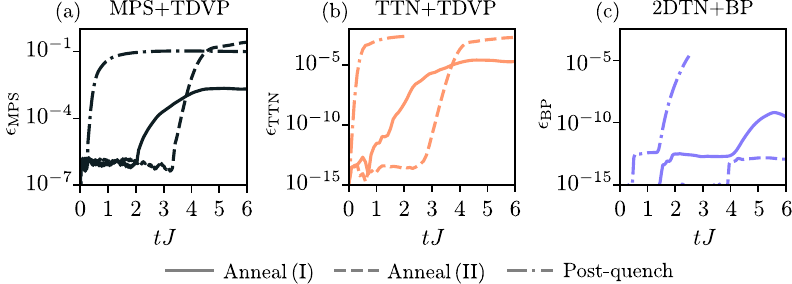}
    \caption{Temporal behavior of the projection errors for tensor network approaches: (a) MPS-TDVP, (b) TTN-TDVP and (c) 2DTN-BP.
    For each case, we consider results for the three different physical regimes,  (i) Annealing I (or slow annealing), (ii) Annealing II (or fast annealing) and (iii) quench dynamics to $h_x/J=2$. In all the cases we fix the system size to $L=10$.}
    \label{fig:convergence_tensornet}
\end{figure*}

\begin{figure}
    \centering
    \includegraphics{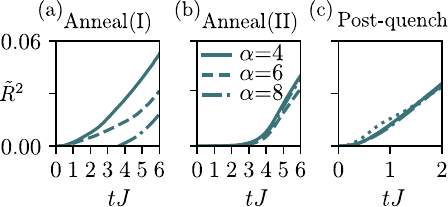}
    \caption{Temporal behavior of the accumulated TDVP error for the NQS-tVMC approach in three different physical scenarios: (a) Anneal~(I), (b) Anneal~(II), (c) post-quench dynamics for $h_x/J=2$. In all the cases, we fix the system size to $L=10$. In addition, we implement our simulations with a two-layer CNN with $\alpha$ and $\alpha -1 $ channels in the two consecutive layers.
    }
    \label{fig:convergence_tvmc}
\end{figure}

\subsection{Results}
\label{sec5_e}

We start discussing the results of the  projection errors for MPS and TTN simulations, as defined in Eqs.~\eqref{eq:mps_error} and ~\eqref{eq:tdvp_err}, respectively; see Fig.~\ref{fig:convergence_tensornet}. For the MPS, the error remains small, of order $\mathcal{O}(10^{-6})$, until the two-site TDVP algorithm reaches the maximum accessible bond dimension within the 40~GB memory limit, at which point the error increases exponentially.
In contrast, the TTN simulations employ the one-site TDVP algorithm, for which the bond dimension is fixed throughout the evolution. In this case, the accumulated error remains small up to the point where the entanglement growth exceeds the representational capacity of the chosen TTN, leading to a rapid rise in the error thereafter.
These trends reflect the characteristic limitations of tensor network methods: as entanglement grows, the restricted expressiveness of the network results in an abrupt loss of accuracy once the effective bond dimension becomes insufficient.
Specifically, for the quench dynamics, where entanglement typically grows linearly in time, this breakdown occurs early in time. For the annealing protocols, the entanglement peaks near the critical point, producing sharp increases in the truncation error before $tJ \simeq 3.0$ for Annealing~(I) and $tJ \simeq 3.8$ for Annealing~(II).
Despite their algorithmic and structural differences, both MPS and TTN display this same qualitative behavior, with the onset of large TDVP errors signaling the failure of the tensor network to efficiently capture the entanglement dynamics.

For the case of 2DTN, the approximate error per Trotter gate during the time evolution, Fig.~\ref{fig:convergence_tensornet}(c), 
also shows a similar trend as the truncation error of MPS and TTN TDVP evolutions, as it starts to increase exponentially once the bond dimension $\chi_\text{2D}$ saturates. We see that this exponential increase is particularly fast for the post-quench case. The annealing results highlight the regimes in which dynamics are well captured by a 2DTN-BP evolution: while for 
Annealing (II) the error remains very low and only starts to increase at later times, for Annealing (I) the error starts to grow much earlier, and even shows a discontinuous behavior at $tJ\approx 5$. The latter can be attributed to numerical instabilities due to a failure of the BP approximation, which are also reflected in the local quantities shown in the (a) panels of Figs.~\ref{fig4}-\ref{fig5}.

Finally, we consider the projection error for the NQS-tVMC approach in Fig.~\ref{fig:convergence_tvmc}; in this case we consider the accumulated error, as defined in Eq.~\eqref{eq:projerror_tdvp}.
In our analysis, we examine results obtained with NQS based on a two-layer CNN, focusing on the dynamics defined by Annealing (I) and (II), as well as the post-quench dynamics to $h_x/J = 2.0$. 
We note that, while for Annealing (I) the quantity $\tilde{R^2}$ systematically decreases as the network size increases, for the other two cases, larger NQS do not reduce $\tilde{R^2}$ beyond a certain threshold time. 
This shows that NQS approaches are suitable for describing quasi-adiabatic annealing processes (such as Annealing (I)), whereas stronger non-adiabatic effects hinder their performance. 
As increasing the ansätze complexity does not improve the results, the approach seems limited by ineffective optimization, not expressivity.
Interestingly, we observe that the time at which $\tilde{R^2}$ abruptly increases in those cases approximately coincides with the point in time where the NQS starts to deviate from the MPS and 2DTN results in the cross-benchmarks discussed in the previous section.

\section{ \label{sec6} Symmetry errors of observables}

Convergence criteria based purely on projection or truncation error can be overly rigid and may fail to reflect the true accuracy of physically relevant quantities. Specifically, using truncation error as the sole measure of convergence is somewhat arbitrary: while it does detect when errors begin to affect the approximate wave function, it does not account for the fact that local observables can remain well-converged even when global properties are not. Setting a hard cutoff on truncation error therefore risks discarding results where local observables are still accurate. For example, in Fig.~\ref{fig:convergence_tensornet} we observe that the truncation error rises exponentially well before $tJ=1$ for post-quench dynamics, meaning that a reasonable truncation-based cutoff would likely be hit very early. Yet, as shown in Figs.~\ref{fig6} and~\ref{fig7}, MPS continues to agree closely with 2DTN results for local observables up to $tJ\approx1.5$, indicating that the truncation error is not a reliable proxy for the convergence of physically relevant quantities.

To address this, we propose an alternative convergence criterion that treats observables individually, using the underlying lattice symmetries as a guiding principle. Importantly, unlike 2DTN and NQS approaches, MPS and TTN ansätze inherently break the full spatial symmetries of a two-dimensional (or higher) lattice because they impose a lower-dimensional tensor network structure. When the bond dimension is too small, this symmetry breaking becomes increasingly pronounced and manifests first in global quantities, before affecting local observables. By explicitly monitoring deviations from lattice symmetries, our approach detects the onset of these errors and provides a physically meaningful, observable-dependent convergence check, avoiding the unnecessary rejection of accurately converged local results.
In particular, we enforce observables conserve quantities that are dependent on geometric symmetries of the Hamiltonian.

\begin{figure*}
    \centering
    \includegraphics{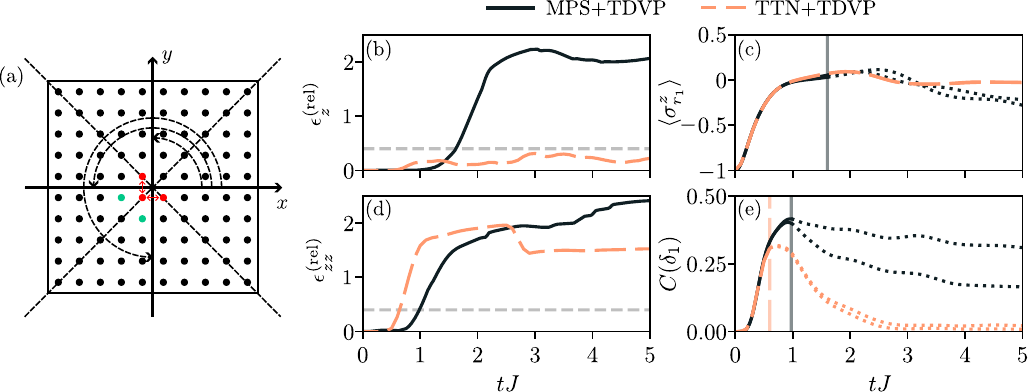}
    \caption{(a) Schematic representation of the $D_4$ symmetry group of the square lattice and the symmetry-equivalent bulk sites considered in panels (b–e): red sites indicate the pairs used to compute connected correlations, while green sites are used for local magnetization. (b) Symmetry error $\epsilon^{(\mathrm{rel})}$ for the local magnetization obtained from MPS (black) and TTN (orange) for post-quench dynamics of a $L=10$ square lattice with $h_x/J = 2.0$. (c) Local magnetization results with the convergence criterion $\epsilon^{(\mathrm{rel})}<0.4$ applied; vertical lines indicate the time beyond which results are no longer considered converged. (d–e) Same as (b–c), but for the connected correlation functions, showing both the growth of the symmetry error and the time at which convergence is lost.}
    \label{fig:sym}
\end{figure*}

Given the uniformity of the external magnetic field and the square grid register, the Hamiltonian exhibits $D_4$ symmetry, depicted in Fig.~\ref{fig:sym}(a), such that rotating sites about angles of $0$, $\pi/2$, $\pi$, and $3\pi/2$ around the grid center, as well as reflections about the $x$, $y$, $x=y$, and $x=-y$ axes (where $x=y=0$ is the grid center) leave the Hamiltonian invariant.
The local observables should also preserve these symmetries, such that observable $\langle O_{i}\rangle$ should equal observable $\langle O_{i'}\rangle$ where $i\rightarrow i'$ is the re-indexing given any one of the 8 symmetry
transformations.
We denote the re-indexed observable $\langle O_{i}^{k}\rangle$ for $k\in1,\dots,7$ for each of the non-trivial transformations (excluding the identity).
From this we define a measure called the symmetry error $\epsilon_i$ that quantifies the maximum degree of $D_4$ asymmetry of a given site $i$ among the local observables as
\begin{align}
    \epsilon_i = \max_{k}(\{|\langle O_i\rangle-\langle O_i^k\rangle|\}_{k=1}^7).
\end{align}
We define the relative symmetry error truncated to a precision of $\xi>0$ for $\langle \bar{O_i}\rangle=(|\langle O_i\rangle|+\sum_{k=1}^7|\langle O_i^k\rangle|)/8$
\begin{align}
    \epsilon^{\mathrm{rel}}_i = 
    \begin{cases}
    \mathrm{if }\langle \bar{O}_i\rangle>\xi,\quad \underset{k}{\mathrm{max}}\left(|\{\langle O_i\rangle-\langle O_i^k\rangle\}_{k=1}^7|/|\langle \bar{O}_i\rangle|\right) \\
    \mathrm{if }\langle \bar{O}_i\rangle\le\xi,\quad \underset{k}{\mathrm{max}}\left(|\{\langle O_i\rangle-\langle O_i^k\rangle\}_{k=1}^7|/\xi\right)
    \end{cases}.
\end{align}
Let us denote the maximum (relative) symmetry error over all observables as $\epsilon^{(\mathrm{rel})}=\underset{i}{\mathrm{max}}(\epsilon_i^{\mathrm{rel}})$.
By asserting that the observables should obey the $D_4$ symmetries, we use $\epsilon^{(\mathrm{rel})}$ to benchmark the quantum simulations as values $\epsilon^{(\mathrm{rel})}>0$ are due to error in the numerical method.

In Fig.~\ref{fig:sym} we present the symmetry-based error metrics for post-quench dynamics of both the local magnetization and the connected correlations, computed for MPS and TTN. We find that these errors remain small over the time window where, for example, MPS and 2DTN agree in the aforementioned plots, consistent with the discussion above. By setting a threshold of $\epsilon^{(\mathrm{rel})}<0.4$, we capture two important features: first, we can detect the point at which observables measured on symmetry-equivalent sites in the bulk (see Fig.~\ref{fig:sym}) begin to deviate from one another; an unphysical result that signals insufficient convergence. Second, this threshold highlights the time at which MPS and TTN results begin to disagree, indicating that at least one of the methods has left the converged regime. This approach therefore provides a single convergence criterion that acts on each observable individually as a necessary condition for the validity of the results, flagging precisely when that observable becomes unreliable. As an illustrative example, we observe that TTN outperforms MPS for local magnetization, likely due to its slightly higher effective dimension, but its connected correlations deteriorate much earlier. This rapid loss of correlation accuracy is a consequence of the TTN structure, where bulk nodes communicate primarily through the top node of the tree, leading to an early breakdown of long-range information.

The convergence criteria proposed in this work acts as a lower bound on the validity of results obtained with MPS and TTN methods. An upper bound could be obtained restricting the same criteria to symmetrically equivalent sites of those that are directly of interest. Furthermore, while the convergence criterion defined here is formulated specifically for square lattices with a globally symmetric Hamiltonian, the underlying idea is entirely general. In practice, the same approach can be extended to any system where the MPS or TTN ansätze breaks exact symmetries of the original problem, by constructing analogous error measures based on those symmetry operations.

\section{Summary of results}
\label{sec7_v2} 

In this Section, we provide a summary of the analyses of capabilities of each numerical approach 
and connect our results with previous works.

The primary limitation of tensor network methods, both MPS and TTN, arises from the growth of entanglement. In the smallest systems we consider, $L=8$, MPS accurately captures the dynamics in all regimes and at all times. As system size increases, and with it the entanglement generated, these methods begin to fail, with long-time post-quench dynamics being the most challenging, followed by annealing II. We identify that these failures manifest as spurious anisotropies in the spatial distributions of magnetization and two-body correlations. Motivated by this observation, we introduce the \textit{symmetry error of observables} as a diagnostic to probe the convergence of both MPS and TTN approaches. Based on the symmetry error we observe that, while a similar performance is seen in both MPS and TTN for local magnetization, in almost all regimes considered a TTN approach starts to struggle to capture the dynamics of correlations at earlier times compared to MPS.

We also investigate the capabilities of two emerging numerical approaches, namely 2DTN and NQS.
We observe that the 2DTN-BP algorithm captures well the short time dynamics in all the regimes and system sizes considered here. One of the advantages of this approach is that the representation is adapted to the 2D lattice geometry, leading to a preservation of its symmetries and
a better scaling of the computational complexity with the system size. For instance, in the large $L$ limit the 2D ansätze can still capture efficiently the fact that at early times the entanglement is expected to play a role only for sites that are close in the 2D plane.

As a drawback of such an approach, we note that the 2DTN algorithm fails abruptly when the BP approximation breaks down significantly, i.e., when the correlations in the lattice deviate strongly from a tree-like behavior. In this situation, encountered in the quasi-adiabatic annealings and later times of post-quench dynamics, the method can become unstable, as signaled by a pathological behavior of local observables (see Figs.~\ref{fig4}-\ref{fig7}), even at the relatively small system size $L=8$. This suggests more accurate contraction methods may be needed to controllably make truncations to the state. This, however, can come at great computational expense and actually lower the effective entanglement that can be captured by the 2DTNs due to bounded computational resources --- resulting in even less accurate results than a BP-based update approach at long times \cite{kim_evidence_2023, king_beyond-classical_2025, tindall_dynamics_2025}.

For the NQS, the results for the quasi-adiabatic dynamics show consistent convergence throughout the entire evolution.
In contrast, for annealing protocols dominated by non-adiabatic effects and for quenches near phase transitions, above a certain timescale, convergence is not achieved even with an increased number of NQS parameters. 
It remains an open problem to identify the reasons behind these limitations. Previous works indicate that this issue is not necessarily related to the expressive power of the NQS, but rather to numerical challenges in computing and solving the tVMC equation of motion using Monte Carlo sampling. Examples of challenges  include (i) the required regularization of the quantum Fisher matrix~\cite{schmitt_quantum_2020} and (ii) the intractably large variance of certain Monte Carlo estimators needed to obtain the tVMC equation of motion~\cite{sinibaldi_unbiasing_2023}.
Finally, it is worth mentioning that in this work we employ dynamics in different local bases (see App~\ref{app:nqs}). In addition to addressing issues related to the representation of the initial state, this approach appears to mitigate some of the difficulties observed in the quench dynamics. The underlying reason for this mitigation, however, remains to be investigated.

\section{Simulating Ising-like Hamiltonians with arrays of Rydberg atoms: experimental considerations}
\label{sec7} 

In this section, we describe how the quantum dynamics that we investigate numerically could also be simulated on a Rydberg-based QPU, where neutral atoms are trapped in programmable tweezer arrays~\cite{browaeys_many-body_2020, saffman_quantum_2010, henriet_quantum_2020}. 

In this platform, the local qubit is encoded in the atomic ground state and a Rydberg state. The Hamiltonian of the QPU, consisting of $N$ neutral atoms trapped in a 2D square lattice geometry (given by interatomic distances $r_{ij}$) can be expressed in terms of Pauli matrices as 
\begin{equation}\label{eq:ham_qpu_2}
\begin{split}
    H_{\text{QPU}} & = \sum_{i<j} J_{ij} \sigma^z_i\sigma^z_j+
    h_x(t) \sum_i \sigma^x_i   + h_z(t)\sum_{i}\sigma^z_i \\
    &+ \sum_i\Delta_{z,i}\left[J_{ij}\right]\sigma^z_i.
\end{split}
\end{equation}
Here $J_{ij} = \frac{C_6}{4 r_{i,j}^6}$ is the fast-decaying atom-atom interaction, whose strength is characterized by the Van der Waals coefficient $C_6$, and leads to a NN interaction $J$ for $r_{ij}$ being NN. We also identify $h_x(t) = \frac{\hbar\Omega(t)}{2}$, being $\Omega(t)$ the Rabi frequency of the ground-Rydberg transition, and $h_z(t)=-\frac{\hbar\delta(t)}{2}+\sum_{i,j}J_{ij}/N$, being $\delta(t)$ the detuning of the laser with respect to the ground-Rydberg transition. Finally,  $\Delta_{z,i}=\sum_{j}J_{ij}-\sum_{l,j}J_{lj}/N$.

When comparing Eqs.~\eqref{eq:H} and \eqref{eq:ham_qpu_2} we observe two differences between $H_\text{QPU}$ and $H$. On the one hand, $H_\text{QPU}$ exhibits a site-dependent longitudinal field $\Delta_{z,i}$, which is non-zero for site-dependent coordination numbers $\sum_{j}J_{ij}$. However, this coordination number is constant in the bulk of the square lattice, so we do not expect an impact in the large system size limit. On the other hand, $H_\text{QPU}$ exhibits long-range interactions beyond NN. Nevertheless, due to their fast $r_{ij}^{-6}$ decay, they are not expected to change the main dynamical phenomena observed under $H_\text{QPU}$ and $H$, but just slightly modify the parameter regimes of both models.

Therefore, it would be interesting to compare the performance of Rydberg-based QPUs in simulating $H_\text{QPU}$ against the numerical methods that we considered in this work to investigate the dynamical properties of $H$. Indeed, for the annealing dynamics the works~\cite{scholl_quantum_2021, scholl_quantum_2021-1, ebadi_quantum_2022} already showed that in this platform one can explore annealing paths with parameter and timescale regimes similar to the ones we considered in this work. For post-quench dynamics, similar 1D experiments were carried out recently~\cite{shaw_benchmarking_2024, choi_preparing_2023}. In the platform, the interaction scale $J$ is typically of a few rad/$\mu$s, and the coherence time $t_{\text{QPU}}$, where unitary dynamics can be well approximated, is usually assumed to be in the few $\mu$s timescale. Thus, timescales of a few units of $tJ= t_{\text{QPU}}[\mu\text{s}]J[\text{rad}/\mu\text{s}]\gtrsim 1$ can be achieved to explore the timescales where, e.g., post-quench dynamics seem challenging to be captured for the numerical methods considered in this work (see Figs.~\ref{fig6}-\ref{fig7}).

\section{\label{sec8} Conclusions and outlook}

In this work, we performed a careful analysis of the capabilities of various state-of-the-art tensor-network based algorithms, as well as a neural quantum state method, for the simulation of the dynamics of different regimes of the 2D TFIM: quasi-adiabatic dynamics, Kibble-Zurek dynamics, and post-quenches dynamics with quenches before and around the quantum phase transition.
We first performed a cross benchmarking of the different approaches to highlight the regimes of convergence and regimes where discrepancies emerge. We then verified that these regimes of convergence and discrepancy are qualitatively in accordance with a projection error analysis. Furthermore, we proposed a new convergence criteria for local observables for methods based on mapping onto an ansätze with a reduced dimensionality compared to the problem.

We found that annealing (I) can be well captured by MPS, TTN and NQS, which describe well the buildup of correlations when entering into the antiferromagnetic phase. In this regime, at later times, we found a 2DTN struggles to efficiently simulate the system due to the strong correlations and small loops in the systems.

In the case of the annealing (II) (Kibble-Zurek regime), meanwhile, we observe an overall agreement between MPS and 2DTN, which seem to capture well the dynamics, whereas TTN and NQS differ when the diabatic effects start to play an important role. The efficacy of a 2DTN approach here is in agreement with similar simulations of quantum annealing in both two and three dimensions \cite{tindall_dynamics_2025}.

For the post-quench dynamics well below the quantum phase transition, all the methods agreed and captured well the oscillations predicted by perturbation theory, as verified by mean-field and semiclassical simulations. However, for the post-quench dynamics around the critical point, all methods struggle to capture the dynamics at long time: NQS falls short in capturing these dynamics even at small system sizes; MPS and TTN capture well the dynamics at small system sizes but the errors increase dramatically with the system sizes; 2DTN also captures the short time dynamics well and the errors are seen to scale much more favorably with system size than other methods.

We then discussed the potential of neutral atom QPUs for simulating these regimes. We first showed that these regimes are  accessible on the current generation of devices. While QPUs will not perform better than classical numerics for quenches in the perturbative regime such as $h_x/J=0.5$, we expect that they should be competitive for regimes $h_x/J=2.0$ and $h_x/J=3.0$.

It is important to emphasize that, while some classical algorithms are very mature such as MPS methods, the field is evolving quickly for other emerging methods. For example, the current limitations of NQS do not necessarily arise from computational complexity related to entanglement, as is the case for tensor-network approaches, but rather from numerical instabilities or sampling issues in the tVMC algorithm. These limitations could be unlocked in the next generation of NQS algorithms.  While we have performed a cross-benchmark of the more well known methods, there are other promising numerical algorithms for dynamics, such as Pauli propagation~\cite{begusic_real-time_2025,broers_exclusive-or_2024,broers_scalable_2025,rudolph_pauli_2025} or influence functional BP~\cite{abanin_colloquium_2019, lerose_influence_2021, park_simulating_2025}.

On the experimental side, we emphasize that, to claim such an advantage, it is essential to carefully address several aspects related to the limitations of the current Rydberg-based analog QPUs. The latter are related to the accuracy of such QPUs in simulating unitary dynamics in the few $\mu s$ duration regime, where it is very challenging for classical numerical methods to deliver a reliable result in some parameter regimes.

First, it is imperative to conduct a careful noise characterization of the QPU to assess whether the corresponding noise levels in current devices enable the accurate capture of both local magnetization and magnetic correlations with an error similar or better than the classical numerical methods. In particular, the characterization of noise at large system sizes, where the numerical methods benchmarked in this work begin to exhibit inaccuracies, can potentially be addressed, given, for example, the recent theoretical evidence showing that for analog QPUs, the errors of physically relevant observables depend only on the noise rate and the simulation time, and not on the system size~\cite{trivedi_quantum_2024, trivedi_noise_2025}.

A second aspect concerns the classical simulability of the dynamics of QPUs in the presence of noise. In this context, studies on the dynamics of digital quantum circuits indicate that noise effects leading to reduced fidelity can make classical simulations more tractable~\cite{ayral_density-matrix_2023,rudolph_pauli_2025}. To the best of out knowledge, it remains unexplored how these results translate to the unitary dynamics realized by analog QPUs based on Rydberg atom arrays. Furthermore, simulations of quantum many-body physics maybe more noise robust in general~\cite{trivedi_noise_2025}. Conducting further research to address this question will be paramount for benchmarking quantum advantage in the future.

Looking forward, continued improvements in the scalability and accuracy of QPUs, particularly in analog platforms such as Rydberg atom arrays, are expected to open new avenues for experimentally exploring and validating quantum dynamics in regimes inaccessible to classical simulation in the near future.

\acknowledgments
We thank Markus Heyl, Lukas Broers, Seiji Yunoki, and Victor Drouin-Touchette for carefully reading the manuscript and providing insightful suggestions. We also thank Antoine Browaeys and Loïc Henriet for fruitful discussions.
Pasqal's team acknowledges funding from the European Union the projects PASQuanS2.1 (HORIZON-CL4-2022-QUANTUM02-SGA, Grant Agreement 101113690). S.B.J. acknowledges support from the Deutsche Forschungsgemeinschaft (DFG, German Research Foundation) under Project No. 277625399-TRR 185 OSCAR (“Open System Control of
Atomic and Photonic Matter”, B4) and under Germany’s Excellence Strategy – Cluster of Excellence Matter and Light for Quantum Computing (ML4Q) EXC 2004/1 – 390534769.
M.S.~and W.K.~were supported through the Helmholtz Initiative and Networking Fund, Grant No. VH-NG-1711. 
J.T. is grateful for ongoing support through the Flatiron Institute, a division of the Simons Foundation.

\appendix

\section{Probes of the Quantum Kibble-Zurek (QKZ) Mechanism}
\label{sec:qkz}

In this supplementary  section, we discuss physical aspects related to the two annealing protocols considered in Sec. \ref{sec4.1}.

According to the QKZ mechanism, during an annealing sweep across a continuous quantum critical point, the rapid divergence of the ground-state correlation length in the vicinity of the transition point can no longer be followed once non-adiabatic effects become significant.
Instead, the system “freezes out” over a characteristic time scale $\hat{t}$ and length scale $\hat{\eta}$, leading to universal scaling laws for excitations and correlations. The QKZ mechanism has been explored in one-dimensional systems~\cite{zhang_observation_2025, keesling_quantum_2019}, but quantitative tests in two-dimensional interacting quantum matter remain comparatively scarce. Recent work by Schmitt \textit{et. al.} in Ref.~\cite{schmitt_quantum_2022} provides a numerical analysis of QKZ scaling in 2D, establishing clear data collapse of correlation functions and excitation energies under appropriate rescaling according to the non-equilibrium scaling hypothesis. 

In our study, we investigate the 2D Ising model on a square lattice under two distinct annealing protocols, Annealing (I) and Annealing (II), in this section we probe whether the QKZ mechanism governs the resulting dynamics. To do so, we replicate the scaling analysis performed in Ref.~\cite{schmitt_quantum_2022} by computing the equal-time spin–spin correlation function close to the critical point during the annealing schedule and rescaling both the correlation distance and correlation amplitude using the theoretically predicted QKZ freeze-out scales. This procedure provides a direct diagnostic of universal KZ behavior: if the annealing schedule places the system within the QKZ regime; meaning the freeze-out length $\hat{\eta}$ remains well below the system size, the rescaled correlation functions obtained at various quench rates must collapse onto a single universal curve.

\begin{figure*}
    \centering
    \includegraphics{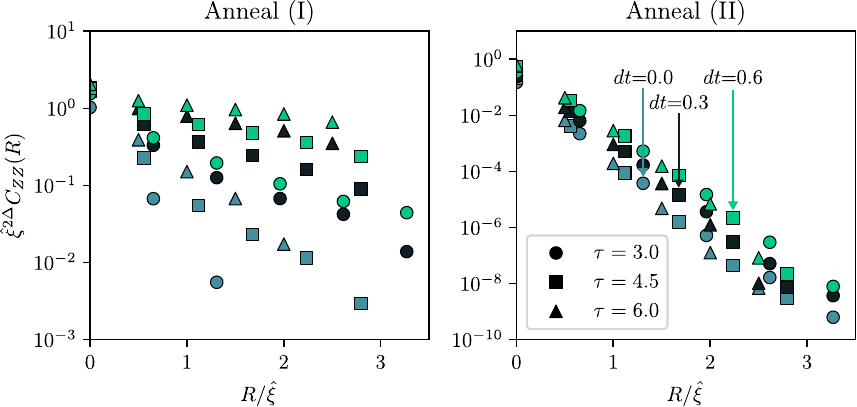}
    \caption{{\bf Connected correlations in the vicinity of crossing points.} The graph shows the scaled correlation function as a function of the scaled distance for three values of the ramp parameter $\tau$, where $\tau$ represents either $t_{\mathrm{fall}}$ or $t_{\mathrm{sweep}}$ in annealing (I) and (II), respectively. Furthermore, $dt = t - t_c$, where $t_c$ is a reference time close to the crossing point, with $t_c J = 3.0$ and $t_c J = 3.8$ for the left and right plots, respectively. 
In this regime, the scaled plots are expected to collapse onto a single scaling function in accordance with the Kibble–Zurek scaling hypothesis. We observe that this is indeed the case for annealing (II), while for annealing (I) we do not observe such a collapse of the results, indicating that the system is in an adiabatic regime.}
    \label{fig_sm_qkz}
\end{figure*}

Our results in Fig.~\ref{fig_sm_qkz} show a sharp distinction between the two annealing schedules. Under Annealing (I), the rescaled correlation functions fail to collapse, indicating that, due to the finite-size gap in the vicinity of the crossing point, non-adiabatic effects do not play a fundamental role and that the dynamics is in the quasi-adiabatic regime. 
In contrast, Annealing (II) produces a robust collapse of the rescaled correlations across a wide range of annealing times. This behavior mirrors the  finite-size scaling results reported in Ref.~\cite{schmitt_quantum_2022} and demonstrates that Annealing (II) traverses the near-critical region at rates that generate universal excitations where $\hat{\eta}/L\ll1$. This findings confirm that only Annealing (II) probes the QKZ regime in our simulations. 

Note, in Fig.~\ref{fig_sm_qkz} we use the theoretically predicted values for rescaling, i.e. $\hat{\xi} \propto\tau^{\nu/(1+z\nu)}$ and $2\Delta = 1+\eta$ with $z=1$, $\nu=0.629971$ and $\eta = 0.036298(2)$. In the work of Schmitt \textit{et. al.} it was found that finite size effects lead to a slight deviation from these values - it is likely that an even stronger collapse could observed in the results of this work with similar considerations.

\section{Additional results}
\label{app:results}

\subsection{Further tests of the Symmetry Convergence Criterion}

In this supplementary section, we provide additional details on the implementation of the symmetry-based convergence criterion introduced in the main text. Our goal is to show how this criterion can be used to regularize numerical results and identify regions of trustworthy dynamics. We focus primarily on MPS, as it is the most widely used method in the literature, examine results with systematically increasing bond dimensions, results obtained from different MPS mappings, and investigate the effect of increasing system size on the timescales over which the dynamics remain converged according to our criterion. Finally, we compare the “converged” results from MPS and TTN with those obtained from benchmark methods, including 2DTN and NQS, to assess the reliability and limitations of the different tensor-network approaches.

\begin{figure}
    \centering
    \includegraphics{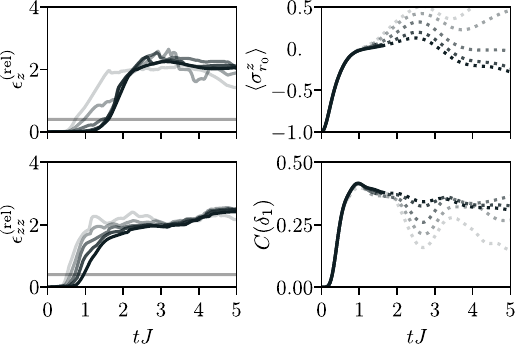}
    \caption{(a1) Symmetry errors for a quench on a $10\times10$ square lattice with $h_x/J=2$, computed for MPS with bond dimensions ranging from 64 to 1024 (powers of two). (a2) Corresponding dynamics of the local observables; the dotted portions of the curves indicate times beyond which the results are no longer considered “converged” according to the symmetry-based criterion. (b1) Symmetry errors for connected correlations with the same set of bond dimensions. (b2) Resulting dynamics for the connected correlations, with dotted segments marking the unconverged regions.}
    \label{fig:fig_sm1}
\end{figure}

In Fig.~\ref{fig:fig_sm1} we show the symmetry errors and corresponding dynamics for both local observables, panels (a1–a2), and connected correlation, panels (b1–b2) as a function of time for increasing MPS bond dimension. As expected, lower bond dimensions lead to earlier growth of asymmetries, indicating a breakdown of the approximate wave function. The convergence criterion effectively identifies the portions of the dynamics where the results begin to deviate as the bond dimension is increased, marking the boundary of the reliably converged regime. Interestingly, the criterion also reveals that connected correlations start to deviate from one another earlier than local observables, highlighting that long-range quantities are more sensitive to the limitations of the MPS representation.

\begin{figure}
    \centering
    \includegraphics{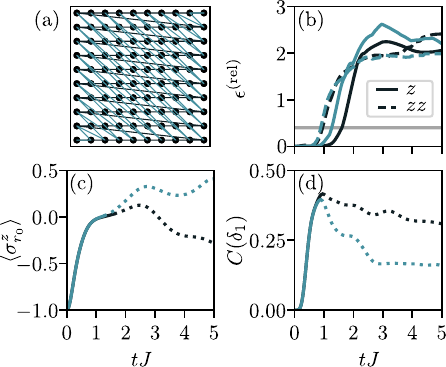}
    \caption{(a) Illustration of two MPS mappings through the $10\times10$ square lattice: horizontal snaking (black) and diagonal snaking (green). (b) Corresponding symmetry errors for both local magnetization and connected correlations for a quench to $h_x/J = 2.0$. (c) Dynamics of the local magnetization and (d) connected correlations as a function of time; dotted points indicate regions where the results are no longer considered converged according to the symmetry-based criterion.}
    \label{fig:fig_sm2}
\end{figure}

In Fig.~\ref{fig:fig_sm2} we compare results obtained from two different MPS mappings of the $10\times10$ square lattice: a horizontal snaking pattern (black) and a diagonal snaking pattern (green). Panel (a) illustrates the two mappings, while panel (b) shows the corresponding symmetry errors for both local magnetization and connected correlations following a quench to $h_x/J = 2.0$. Panels (c) and (d) display the time evolution of these observables, with dotted points indicating regions where the results are no longer considered converged according to the symmetry-based criterion.

It is well known that the choice of MPS mapping plays a critical role in accurately capturing long-range correlations. Optimal mappings are typically chosen to preserve locality: nodes that are connected in the MPS should correspond to physically nearby sites on the lattice. Different mapping choices can therefore lead to quantitatively different results, particularly for observables sensitive to long-range correlations. It is also interesting to note that different mappings will break or preserve different symmetries from the $D_4$ subgroup, which could further influence the growth of symmetry errors and the apparent convergence of the simulation.

This sensitivity provides a natural test for the convergence criterion. By comparing results obtained from different mappings, we can assess whether the criterion reliably identifies portions of the dynamics where the approximate MPS wave function ceases to be accurate. As shown in panels (c) and (d) of Fig.~\ref{fig:fig_sm2}, for both local magnetization and connected correlations, the divergence between the two mappings coincides with the symmetry-error criterion flagging the results as unconverged. This demonstrates that the convergence criterion effectively captures the breakdown of the approximation and identifies the time windows over which the dynamics can be trusted.

\begin{figure}
    \centering
    \includegraphics{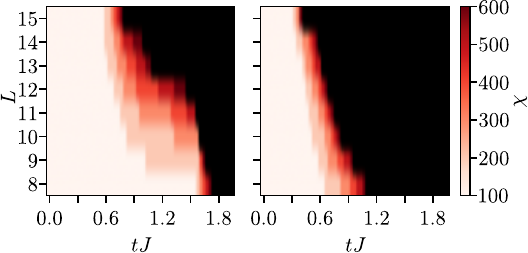}
    \caption{The bond dimension needed to achieve convergence is shown for various system sizes, with a maximum of 600 due to memory constraints for the $14\times14$ lattice. Note, the black region shows where this maximum bond dimension was not enough to achieve convergence. Panel (a) shows results for local magnetization, while panel (b) displays connected correlations.}
    \label{fig:fig_sm3}
\end{figure}

In Fig.~\ref{fig:fig_sm3} we show the bond dimension required to achieve convergence for both local magnetization (panel a) and connected correlations (panel b) as a function of system size and simulation time. For short-time dynamics, a relatively small bond dimension is sufficient, as the amount of entanglement generated during these early times remains modest. However, as the simulation progresses to longer times, the required bond dimension increases to accurately capture the growing entanglement. Similarly, larger system sizes demand higher bond dimensions for the same simulation time, reflecting the known result that capturing long-range correlations in extended systems becomes increasingly difficult. Notably, we see directly that, to converge to the same time and system size, connected correlations require higher bond dimensions than local magnetization, highlighting their greater sensitivity to the limitations of the MPS representation.

\subsection{Full annealing results}

\begin{figure}
    \centering
    \includegraphics{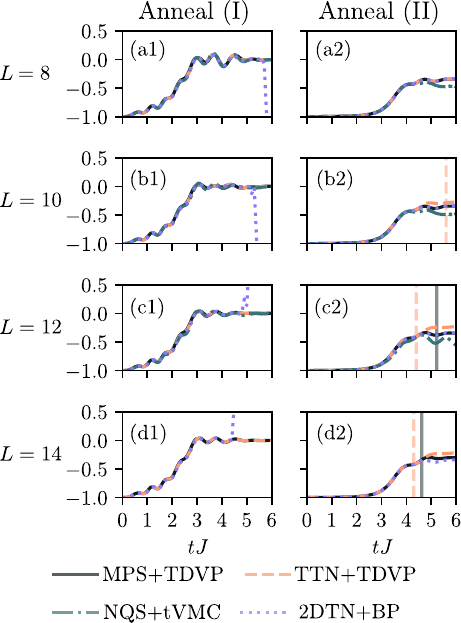}
    \caption{Time evolution of the local magnetization, $\langle \sigma_{r_0}^z \rangle$, for anneal sweeps (I) and (II) (corresponding to panel a1 and b1 in Fig.~\ref{fig4}) shown for all system sizes studied. Results are obtained with MPS, TTN, NQS, and 2DTN. Vertical lines indicate the times up to which MPS and TTN results are considered converged according to the symmetry-based convergence criterion.}
    \label{fig:fig_sm4}
\end{figure}

\begin{figure}
    \centering
    \includegraphics{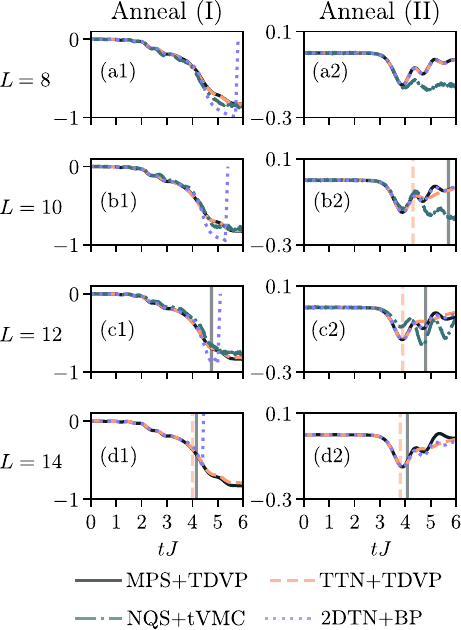}
    \caption{Time evolution of the nearest-neighbor connected correlation, $C(\delta_1)$, for anneal sweeps (I) and (II) (corresponding to panel a1 and b1 in Fig.~\ref{fig5}) shown for all system sizes studied. Results are obtained with MPS, TTN, NQS, and 2DTN. Vertical lines indicate the times up to which MPS and TTN results are considered converged according to the symmetry-based convergence criterion.}
    \label{fig:fig_sm5}
\end{figure}

In Figs.~\ref{fig:fig_sm4} and~\ref{fig:fig_sm5} we present the full set of annealing results analyzed in this work for all system sizes studied. Figure~\ref{fig:fig_sm4} shows the time evolution of the local magnetization, $\langle \sigma_{r_0}^z \rangle$, for annealing sweeps (I) and (II) (corresponding to panels a1 and b1 in Fig.~\ref{fig4} of the main text), while Fig.~\ref{fig:fig_sm5} displays the nearest-neighbor connected correlations, $C(\delta_1)$, for the same sweeps (corresponding to panels a1 and b1 in Fig.~\ref{fig5}). Results are obtained with MPS, TTN, NQS, and 2DTN.

As the annealing protocol is also defined on a square lattice and the Hamiltonian respects the $D_4$ symmetry, the same symmetry-based convergence criterion used for the quench dynamics can be applied here. Deviations in the local observables and connected correlations along symmetry-equivalent sites indicate the onset of unphysical asymmetries, allowing us to identify the time windows where MPS and TTN simulations remain reliable. Thus, vertical lines indicate the times up to which MPS and TTN results are considered converged according to the symmetry-based convergence criterion.

These figures demonstrate that the convergence criterion successfully identifies the times at which MPS and TTN begin to deviate from each other as well as from other methods in most cases, providing a reliable measure of the time window over which each method produces trustworthy dynamics. Across all system sizes, the criterion captures both early-time agreement and the onset of deviations in long-time dynamics, validating its effectiveness as a practical tool for benchmarking tensor-network simulations.

\subsection{Full post-quench dynamics results}

\begin{figure*}
    \centering
    \includegraphics{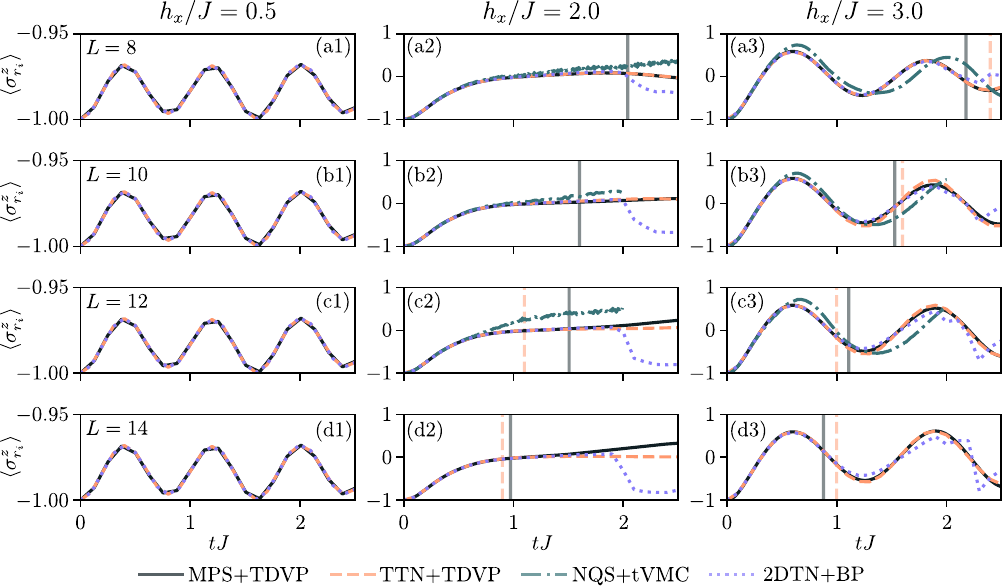}
    \caption{Time evolution of the local magnetization, $\langle \sigma_{r_0}^z \rangle$, for post-quench dynamics corresponding to panels (a1), (b1), and (c1) in Fig.~\ref{fig6} of the main text, shown here for all system sizes studied. The three columns correspond to quench strengths $h_x/J = 0.5$, $h_x/J = 2.0$, and $h_x/J = 3.0$. Results are obtained with MPS, TTN, NQS, and 2DTN. Vertical lines indicate the times up to which MPS and TTN results are considered converged according to the symmetry-based convergence criterion.}
    \label{fig:fig_sm6}
\end{figure*}

\begin{figure*}
    \centering
    \includegraphics{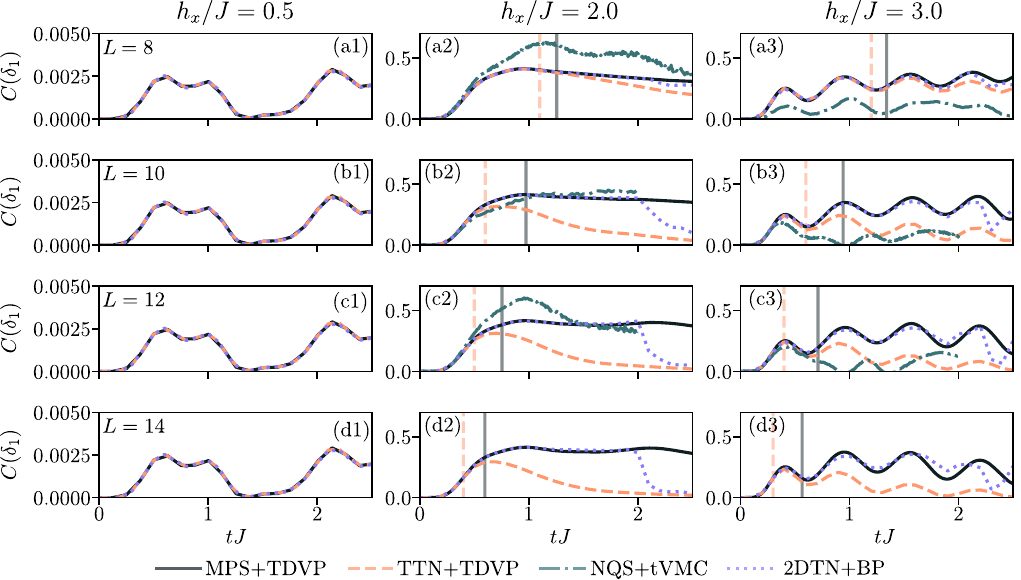}
    \caption{Time evolution of the nearest-neighbor connected correlation, $C(\delta_1)$, for post-quench dynamics corresponding to panels (a1), (b1), and (c1) in Fig.~\ref{fig7} of the main text, shown here for all system sizes studied. The three columns correspond to quench strengths $h_x/J = 0.5$, $h_x/J = 2.0$, and $h_x/J = 3.0$. Results are obtained with MPS, TTN, NQS, and 2DTN. Vertical lines indicate the times up to which MPS and TTN results are considered converged according to the symmetry-based convergence criterion.}
    \label{fig:fig_sm7}
\end{figure*}

In Figs.~\ref{fig:fig_sm6} and~\ref{fig:fig_sm7} we present the full set of post-quench dynamics results analyzed in this work for all system sizes studied. Figure~\ref{fig:fig_sm6} shows the time evolution of the local magnetization, $\langle \sigma_{r_0}^z \rangle$, for quench strengths $h_x/J = 0.5$, $2.0$, and $3.0$ (corresponding to panels a1, b1, and c1 in Fig.~\ref{fig6} of the main text). Figure~\ref{fig:fig_sm7} displays the nearest-neighbor connected correlations, $C(\delta_1)$, for the same quench strengths (corresponding to panels a1, b1, and c1 in Fig.~\ref{fig7} of the main text). Results are obtained with MPS, TTN, NQS, and 2DTN. Vertical lines indicate the times up to which MPS and TTN results are considered converged according to the symmetry-based convergence criterion.

These figures demonstrate that the convergence criterion reliably identifies the times at which MPS and TTN begin to deviate from each other as well as from other numerical methods in most cases. Across all system sizes and quench strengths, the criterion captures both early-time agreement and the onset of deviations in long-time dynamics, validating its effectiveness as a practical tool for benchmarking tensor-network simulations of quench dynamics.


\section{TDVP error for tree tensor networks} \label{app:tdvp_err}

In this section we provide a derivation for the TDVP error~\eqref{eq:tdvp_error_ttn} used for quantifying the accuracy of the TTN simulations. Starting from eq.~\eqref{eq:tdvp_err}, we rewrite the evolved variational state using the tangent-space projector $\hat{P}$

\begin{equation}
    \ket{\psi_{\vec{\theta}(t + \delta t)}} = e^{-i\hat P \hat H \delta t}\ket{\psi_{\vec{\theta}(t)}}\ ,
\end{equation}

which is the form typically used in formulations of the TDVP equation for tensor network states. Following the derivation from~\cite{schmitt_quantum_2020}, consisting of an expansion of the exponential up to second order in $\delta t$, and identifying $\mathcal{E}=-i\hat H \delta t$ and $R=-i\hat P \hat H \delta t$, we arrive at eq.~\eqref{eq:tdvp_error_ttn} from the main text.

Note, that the quantities appearing in the equation can be calculated efficiently using the already existing, effective environments of the TDVP algorithm.

\section{Neural quantum states}
\label{app:nqs}

In this Section, we describe some important  details of the NQS-tVMC simulations considered in this work.

We parametrize the wave function $\psi_{\vec{\theta}}(s)$, where $\vec{\theta}$ is a set of variational parameters parametrised by  NQS architectures, which consist
of a two-layer Convolutional Neural Networks with $\alpha$ and $\alpha -1 $ channels in the two consecutive layers.  The equation of motion of such variational parameters is given by 
\begin{equation}
    S_{k,k'} \dot{\theta}_{k'}=-iF_k\ ,
    \label{eq:tdvp}
\end{equation}
where $\dot{\theta}_{k'}$ is the time derivative, 
\begin{equation}
\begin{split}
S_{k,k'}=\braket{\partial_{\theta_k}\psi_{\vec\theta}|\partial_{\theta_{k'}}\psi_{\vec\theta}}
-\braket{\partial_{\theta_k}\psi_{\vec\theta}|\psi_{\vec\theta}}\braket{\psi_{\vec\theta}|\partial_{\theta_{k'}}\psi_{\vec\theta}}
\label{Skk}
\end{split}
\end{equation}
is the quantum Fisher matrix and 
\begin{equation}
\begin{split}
F_{k}=\braket{\partial_{\theta_k}\psi_{\vec\theta}|\hat H|\psi_{\vec\theta}}-\braket{\partial_{\theta_k}\psi_{\vec\theta}|\psi_{\vec\theta}}\braket{\psi_{\vec\theta}|\hat H|\psi_{\vec\theta}}.
\label{Fk}
\end{split}
\end{equation}
Both $S_{k,k'}$ and $F_k$ can be estimated via Monte-Carlo sampling of the Born distribution $|\psi_{\vec\theta}(s)|^2$.

The time evolution of the variational parameters $\theta$ is performed
by numerically solving Eq. \ref{eq:tdvp} with the
second-order Heun method 
with a fixed time interval of $\delta_t J =  10^{-2}$.

Furthermore, an important technical detail is that the $S$ matrix turns out to be ill-conditioned (i.e., it can contain an eigenvalue spectrum that spans all numerical orders of magnitude). Therefore, the time evolution via tVMC requires suitable regularization schemes when inverting the $S$ matrix to solve  Eq.\eqref{eq:tdvp}. In particular, we consider the scheme proposed in ref.~\cite{schmitt_quantum_2020}, which is based on a targeted elimination of noisy contributions of singular values of $S$.

\subsubsection{NQS-tVMC simulations with different local basis}
\label{app:localbasisNQS}
A second important detail is that the simulations of the dynamics of the initial-time product state in the computational basis (as in Eq.\ref{initState}) pose issues for tVMC simulations. First, simply because the $S$-matrix vanishes for such a state. Second, the variational states whose probability distribution ($|\psi_{\vec\theta}(s)|^2$) is equal to zero for many basis states, but yield finite contributions to the gradients,  is know to bias estimators of certain observables (e.g., as the ones necessary to access $F$), which can demand an exponentially large number of samples~\cite{sinibaldi_unbiasing_2023}.

To circumvent such issues, we consider simulations performed in a rotated basis. In order to implement such basis rotations in our simulations, we consider a unitary transformation of the Hamiltonian,
$ \tilde{H}(\gamma) = \hat{U}(\gamma)\, H\, \hat{U}^{\dagger}(\gamma) $,
where $\gamma$ is the parameter of the unitary transformation.
Particularly, in this work, we consider simulations to a more specific choice of local basis defined by the unitary transformation: 
\begin{equation}
\hat{U}(\gamma) = \prod_{k=1}^N e^{-i \sigma_k^y \gamma_k/2},    
\end{equation}
we consider a uniform $\gamma_k = \gamma$.

The time dependence of an observable $A$ can then be written as
\begin{equation}
 \begin{split} 
 \bra{\psi(0)} e^{iHt} \hat{A} e^{-iHt} \ket{\psi(0)} = \\ \bra{\tilde{\psi}(0)} e^{i\tilde{H}t} (\hat{U} \hat{A} \hat{U}^{\dagger}) e^{-i\tilde{H}t} \ket{\tilde{\psi}(0)}.
  \end{split} 
  \end{equation}
Our strategy is therefore as follows: instead of simulating the dynamics generated by the original Hamiltonian $H$, we simulate the dynamics of the transformed Hamiltonian $\tilde{H}$, with the initial state ${\ket{\tilde{\psi}(0)}}$ being the ground state of Hamiltonian $\tilde{H}(\gamma)$ with $h_x = 0$.

For all the results presented in Sec.~\ref{sec2.1} for the quantum annealing protocols we consider $\frac{\gamma}{\pi} = 0.1$, while in Sec.~\ref{sec2.2} for the post-quench dynamics we consider $\frac{\gamma}{\pi} = 0.1$ for $h_x/J=2.0$ and $\frac{\gamma}{\pi} = 0.3$ for $h_x/J = 0.3$.

\section{Two-dimensional tensor networks with belief propagation}
\label{app:2dtn}
\begin{figure}
    \centering
    \includegraphics[width=\columnwidth]{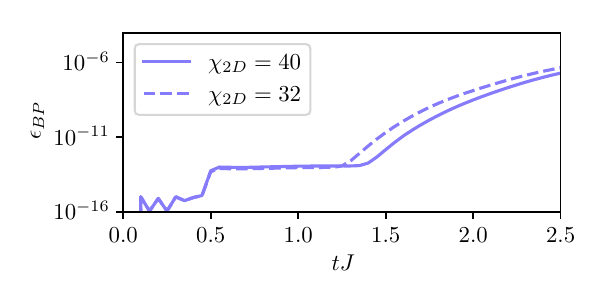}
    \caption{Comparison of the growth of the truncation error in the 2DTN-BP method under two different bond dimensions $\chi_{2D}$. Here we considered the post-quench dynamics for $h_x/J=2.0$ and $L=10$.}
    \label{fig:BP_trunc_scaling}
\end{figure}

In this section, we discuss the effect of changing the bond dimension used in  the 2DTN-BP method. In Fig.~\ref{fig:BP_trunc_scaling}, we illustrate the behavior for the particular case of post-quench dynamics at $h_x/J=2.0$ and $L=10$, for which we show the difference in the truncation error for $\chi_{2D}=40$ (used to obtain the results of the main text) and a lower $\chi_{2D}=32$. In terms of peak RAM during these Trotter evolutions, the simulation at $\chi_{2D}=32$ used $\approx 12$ GB, while the one with $\chi_{2D}=40$ used $\approx 24$ GB. Despite this increase in the computational resources, we observe that increasing the bond dimension only slightly shifts the time at which the truncation errors starts to grow exponentially. 

\section{Simulation Runtimes}
\label{app:runtimes}

In this section, we summarize the computational runtimes for the numerical methods used in this work. The simulations were performed for two annealing schedules, Anneal~(I) and Anneal~(II), followed by post-quench dynamics with $h_x/J=2.0$. The runtimes reported here correspond to the average time required to complete a single time-step for a given simulation, including all stages of the calculation, see Table.~\ref{tab:runtimes}.

\begin{table*}
    \centering
    \begin{tabular}{|c|c|c|c|c|}
        \hline
        Lattice Size & MPS-TDVP~\cite{paeckel_time-evolution_2019, stoudenmire_studying_2012} & TTN-TDVP~\cite{krinitsin_time_2025} & NQS-tVMC \cite{schmitt_jvmc_2022} & 2DTN-BP / bMPS \cite{tindall_gauging_2023, lubasch_algorithms_2014} \\
        \hline
        8 $\times$ 8   & 10.4 & 4.5 &     10.2             &       3.3 / 3.6             \\
        10 $\times$ 10 & 10.6 & 7.0 &       11.8           &   5.7 / 7.6                \\
        12 $\times$ 12 & 8.9 & 8.2 &       26.9           &    8.2  / 11.7              \\
        14 $\times$ 14 & 7.3 & 10.0 &       N/A           &       11.5/ 20.3             \\
        \hline
        \hline
        FLOPS & $O(N^{3/2} \chi^3)$ & $O(N \log(N) \chi^4)$ &    $O( N_{\mathrm{MC}}  \max(N^2,N_v) \ N_v)$  & $O(\chi_{2D}^{5}N) / O(\chi_{2D}'^{8}N)$
        \\
        Memory & $O( N^{3/2} \chi^2)$ & $O(N \log(N) \chi^3)$ &   $O(N_v^2)$   & $O(\chi_{2D}^{4}N) / O(\chi_{2D}'^{ 6}N)$
        \\
        \hline
        
    \end{tabular}
    \caption{Average runtime (in hours) for each method to simulate dynamics for a period of $tJ = 1$ using a single NVIDIA A100 GPU. Note that this runtime includes the time-evolution of the quantum state, as well as the computation of the local magnetization and spin-spin correlations. For 2DTN, we report separately the runtime of the time evolution (performed on CPU) and the computation of observables at ten intervals of width $\Delta tJ = 0.1$ using bMPS (on GPU). We note that GPU-based time evolution is expected to provide substantial speedups---around an order of magnitude ($\sim 10\times$)---at the larger bond dimensions. We also assume the boundary-MPS dimension is comparable to that of the state, $r_{2D} \sim \mathcal{O}(\chi'_{2D})$. For NQS-tVMC, we assume an approximately linear scaling of runtime with the number of GPUs~\cite{schmitt_jvmc_2022}, and therefore list the actual runtime multiplied by the number of GPUs used. We also report the FLOP and memory scaling for each method, where $\chi$ denotes the TN bond dimension. For NQS, $N_{\mathrm{MC}}$ is  the number of Monte Carlo samples, and $N_v$ the number of variational parameters (which in the cases considered here $N_v \sim N$, $N$ being the number of qubits). In MPS simulations, the bond dimension was chosen such that the 40-GB memory capacity of an NVIDIA A100 GPU was saturated; because the maximum feasible bond dimension depends strongly on system size, preventing reliable conclusions about how the runtime scales with system size.}
    \label{tab:runtimes}
\end{table*}

In the MPS simulations performed, the bond dimension was dynamically adjusted to saturate the full 40~GB memory capacity of an NVIDIA A100 GPU. Because the maximum feasible bond dimension decreases with increasing system size, this leads to system-size–dependent truncation thresholds. For instance, when simulating a system of size $L=8$, the GPU memory allowed us to reach a bond dimension of 1600. In contrast, for $L=14$ the same memory constraint limited the bond dimension to 700. As a result, the runtime cannot be meaningfully compared across system sizes, since each simulation effectively uses a different truncation regime.

\section{Mean-field and semiclassical solutions}
\label{app:approx_sol}

\begin{figure*}
    \centering
\includegraphics[width=\textwidth]{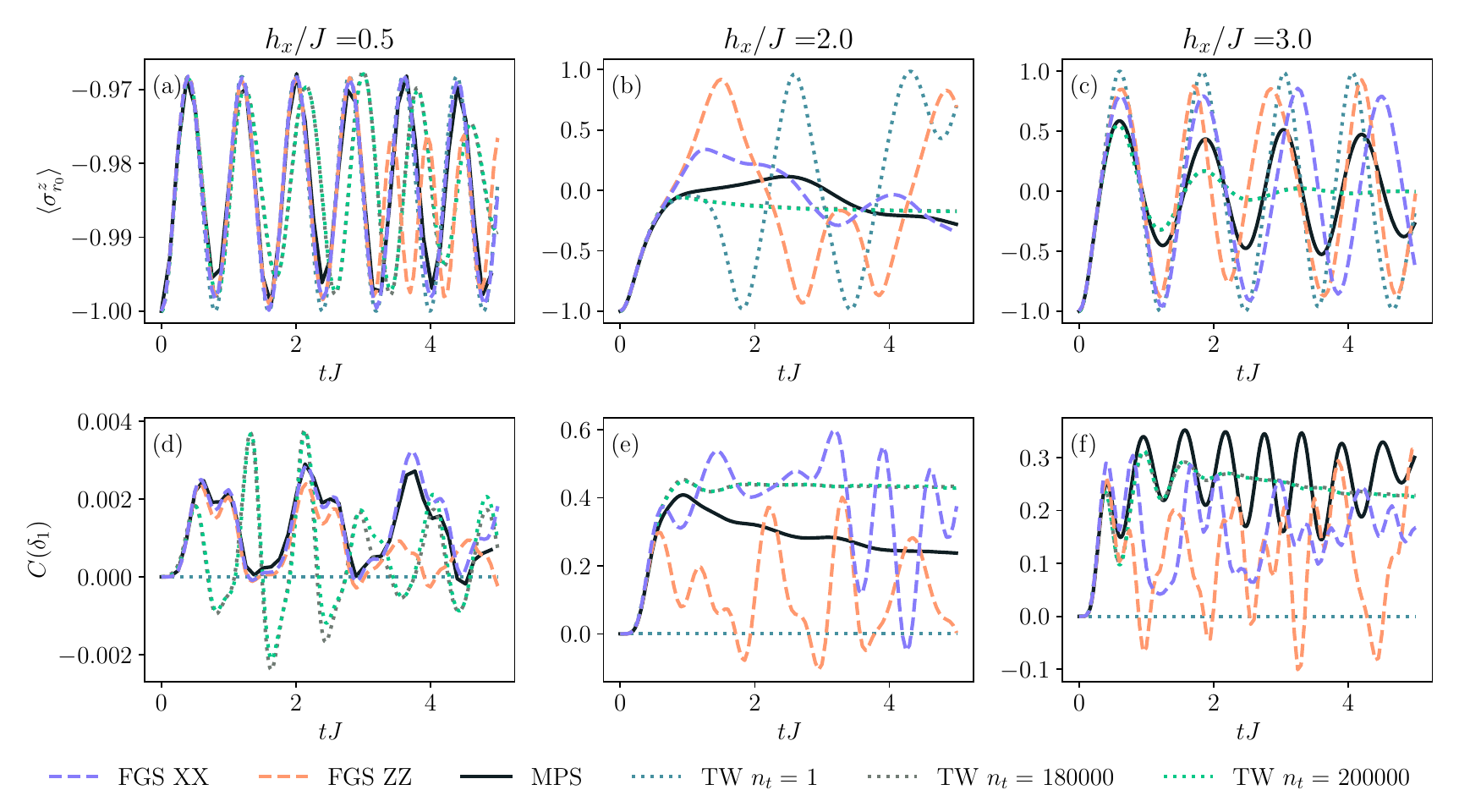}
    \caption{Mean-field and semi-classical results: Real-time dynamics after a global quench from the state $|00\dots 0\rangle$ for a $(10\times 10)$ square lattice. Columns correspond to different transversal field strengths $h_x/J\in\{0.5, 2.0, 3.0\}$. The top (bottom) row  displays the central site magnetization $\langle\sigma^z_{r_0}\rangle$ (central site connected-correlation $C(\delta_1)$). Black solid line denotes the MPS results. We present results for the PV-FGS ansätze in the basis where the interaction is giving by $\sigma^z_k\sigma^z_l$ ($\sigma^x_k\sigma^x_l$) in dashed orange (purple). Results from the TW simulations using different numbers of samples $n_t$ by dotted lines of shades of green, with $n_t=1$ corresponding to the spin mean-field (SMF) solution.}
      \label{fig:mf_sc_res}
\end{figure*}

While the main part of this work focused on strongly correlated methods, these methods quickly become inefficient in describing large system sizes in higher dimensions. For such systems, the pool of available numerical methods which can still be used to simulate the underlying spin dynamics while capturing some form of entanglement or correlation is much more limited. In light of rapid improvements in experimental control of quantum systems beyond 100 particles, one has to resort to computationally less demanding many-body methods that are able to deal with large system sizes. However, such computational methods are often restricted to families of quantum states which can only capture a limited amount of entanglement and typically only work well on short time scales. 

In the following, we consider three methods that can be applied to study the dynamics of large spin-1/2 systems: We will consider (i) a spin mean-field (SMF) theory, (ii) a fermionic mean-field theory based on parity-violating fermionic Gaussian states (PV-FGS), and (iii) a method based on the discrete truncated Wigner approximation (TW). While (i) can be considered as a purely classical mean-field solution incapable of capturing correlations or entanglement, (ii) adds quadratic corrections to its mean-field solution and is able to introduce  entanglement in spin systems~\cite{peschel_reduced_2009,kaicher_mean-field_2023}. Method (iii) describes a semiclassical theory which is able to accurately describe the  short-time non-equilibrium dynamics of large spin systems~\cite{schachenmayer_many-body_2015}. 

\paragraph{Spin mean-field theory:}
A system of non-interacting spins can be described by a  spin-1/2 product state 
\begin{align}
    \ket{\Psi}  =  \bigotimes_{j=1}^N \left(\alpha_j\ket{0}_j + \beta_j\ket{1}_j\right),\label{n109}
\end{align}
where $\alpha_j\in \mathds R$ and $\beta_j\in\mathds C$, with $
|\alpha_j|^2 + |\beta_j|^2=1$. The mean-field approximation is given by  $\sigma^z_i \sigma^z_j\approx \sigma^z_i \braket{\sigma^z_j} +  \braket{\sigma^z_i}\sigma^z_j-\braket{\sigma^z_i}\braket{\sigma^z_j}$ which leads to en effective self-consistent mean-field potential~\cite{weiss_hypothese_1907}. This approximation works usually well if quadratic fluctuations around the mean value of the spin variables are small. Using a time-dependent variational principle, real-time equations of motion can be derived either for the variational wave function parameters $\alpha_j$ and $\beta_j$~\cite{mauron_predicting_2025}, or, equivalently, for the spin observables directly~\cite{schachenmayer_many-body_2015}.

\paragraph{Fermionic mean-field theory:}
Similar to Eq.~\eqref{n109}, we consider parity-violating fermionic Gaussian states (PV-FGS)~\cite{seifi_mirjafarlou_generalization_2024}
\begin{align}
    \ket{\Psi}  =& \exp\left( \frac{1}{2} \begin{pmatrix} \mathbf c^\dag & \mathbf c \end{pmatrix} \mathbf M \begin{pmatrix}
        \mathbf c \\ \mathbf c^\dag
    \end{pmatrix} +\mathbf u^\dag \mathbf c^\dag +\mathbf v^T\mathbf c \right)\ket{0}^{\otimes N},\label{npv_fgs_36}
\end{align}
where $\hat{\mathbf c}=(\hat c_1,\hat c_2,\dots,\hat c_{N})^T$ is a vector of fermionic creation operators, $\mathbf M$ is a suitable matrix, $\mathbf u, \mathbf v$ are vectors, chosen so that the argument of the exponent is anti-Hermitian. 
While previous fermionic mean-field methods have focused on parity preserving FGS, i.e. states where $u_j=v_j=0$ for all $j\in [1,
\dots 2N]$, which can be efficiently described using the method outlined in Ref.~\cite{shi_variational_2018}, studying a spin Hamiltonian that maps to a PV fermionic Hamiltonian, or quenches from an initial state which cannot be described by a parity-preserving FGS require the study of the more complex PV-FGS ansätze~\eqref{npv_fgs_36}, where i.g. $u_j,v_j\neq 0$. A comprehensive account of the method will be given in a forthcoming publication~\cite{KaicherInPrep}.

\paragraph{Discrete truncated Wigner approximation:}
TW formulates spin dynamics in discrete phase space~\cite{schachenmayer_many-body_2015}.
Expectation values are written as
\(
\braket{\hat O(t)}=\sum_{\alpha} w_\alpha\, O^W_\alpha(t),
\)
where \(\alpha\) indexes the \(4^N\) phase points, \(w_\alpha\) is the (time-independent) discrete Wigner function of the initial state, and \(O^W_\alpha(t)\) is the Weyl symbol evolved in time. In practice one samples \(\alpha\) from \(w_\alpha\) and propagates classical spin variables \(\mathbf s_i\) with the Hamiltonian \(H_C(\{\mathbf s\})\) obtained by replacing \(\hat\sigma_i^\mu\!\to s_i^\mu\):
\begin{equation}
\dot s_i^\alpha
= \{s_i^\alpha,H_C\}
= 2\sum_{\beta\gamma}\varepsilon_{\alpha\beta\gamma}\, s_i^\gamma\,\frac{\partial H_C}{\partial s_i^\beta}.
\label{eomcl_short}
\end{equation}
Averages over \(n_t\) samples yield statistical errors \(\sim n_t^{-1/2}\), where $n_t$ is the number of samples, largely independent of \(N\). 

TW is able to capture certain correlations over a finite time provided that the initial state can be sampled with classical trajectories. An important class of such initial states are factorizable spin states. TW has been used and verified to describe correlations in spin systems with short and long-range interactions~\cite{perlin_spin_2020}. The true advantage of TW is its modest computational cost in comparison to other more strongly correlated methods or advanced mean-field theories such as those based on FGS described in the previous paragraph. In the single-sample, no-fluctuation limit TW reduces to SMF.

\paragraph{Results:} Figure~\ref{fig:mf_sc_res} summarizes the results from the mean-field and semiclassical methods introduced above. We compare the results for a $(10\times 10)$ square lattice only against MPS data in order to avoid clutter. The MPS data fulfilling the convergence criteria discussed in the main text are indicated by hollow circles. We compute the single-site magnetization $\langle\sigma^z_{r_0}\rangle$ in the top row and central site connected-correlation $C(\delta_1)$ in the bottom row for the regimes $h_x/J\in \{0.5, 2, 3\}$ in the respective columns of Figs.(a)-(f). For all mean-field and semiclassical methods we solve the equations of motion using a fourth order Runge-Kutta method at step size $\Delta t=0.001$. For $h_x/J\in \{0.5, 3\}$ we used $n_t=180000$ and $200000$ samples, while for $h_x/J=2.0$ we used $n_t=300000$ and $400000$ samples for the TW method (`TW'). We use the PV-FGS for two different basis of the Ising Hamiltonian, i) the original basis where all interactions are described by  $\sigma^z_k\sigma^z_l$-interactions (`FGS') and ii) a rotated basis where the interactions are described by $\sigma^x_k\sigma^x_l$ (`FGS XX'). In the ZZ and XX bases, the initial states are given by $\ket{00\dots 0}$ and $2^{-N/2}(\ket{0}+\ket{1})^{\otimes N}$, respectively. Overall, we find the PV-FGS XX results outperform the results PV-FGS ZZ in all regimes.  The spin-mean field solution (`SMF') fails to describe the qualitative behavior of MPS across all regimes. PV-FGS XX is able to capture the behavior of both observables in the low-field regime $h_x/J=0.5$, but fails to describe the high-field regime $h_x/J=3.0$. In contrast, the initial behavior of TW accurately captures the initial dynamics in the high-field regime, before converging to a classical equilibrium state. However, TW completely fails in describing the connected correlation functions in the low-field limit. All mean-field and semiclassical methods studied here fail to describe the dynamics of the considered observables  at $h_x/J=2.0$ already after the first oscillation. 

\bibliography{quanta-bib}

\end{document}